\documentclass[10pt,twocolumn,floatfix,preprintnumbers,amsmath,amssymb]{revtex4}

\usepackage{graphicx}
\usepackage{dcolumn}%
\usepackage{bm}%

\usepackage{amsmath}
\usepackage{amssymb}
\usepackage{amsfonts}

\usepackage{xifthen}
\usepackage{mathbbol}
\usepackage{subfigure}

\usepackage[a4paper,CJKbookmarks,bookmarks=true,bookmarksopen=true]{hyperref}
\hypersetup{
    a4paper,
    pdftitle={Optimal excitation of two dimensional Holmboe instabilities},
    pdfauthor={Navid C. Constantinou and Petros J. Ioannou},
    pdfkeywords={},
    bookmarksnumbered,
    pagebackref=true,
    breaklinks=true,
    urlcolor=blue,
    bookmarks=true,
    colorlinks=true,
    linkcolor=red,
    citecolor=blue,          
        }

\newcommand{\be}{\begin{equation}}
\newcommand{\ee}{\end{equation}}
\newcommand{\bdm}{\begin{equation*}}
\newcommand{\edm}{\end{equation*}}
\newcommand{\bea}{\begin{eqnarray}}
\newcommand{\eea}{\end{eqnarray}}

\newcommand{\partialf}[2]
{
 \ifthenelse{\equal{#1}{}}{\frac{\partial}{\partial #2}}{\frac{\partial #1}{\partial #2}}
}

\newcommand{\sech}{\mathop{\mathrm{sech}}}

\renewcommand{\(}{\left(}
\renewcommand{\)}{\right)}
\renewcommand{\[}{\left[}
\renewcommand{\]}{\right]}

\renewcommand{\L}{\mathbb{L}}

\renewcommand{\a}{\alpha}

\renewcommand{\r}{\varrho}

\renewcommand{\u}{\mathbf{u}}
\renewcommand{\v}{\mathbf{v}}

\newcommand{\1}{\mathbb{1}}
\newcommand{\A}{\mathbb{A}}

\newcommand{\M}{\mathbb{M}}

\newcommand{\R}{\mathbb{R}}

\newcommand{\D}{\Delta}
\newcommand{\Df}{\text{D}}
\newcommand{\df}{\text{d}}

\newcommand{\F}{\mathbb{F}}

\newcommand{\K}{\mathbb{K}}

\newcommand{\s}{\sigma}

\newcommand{\fb}{\mathbf{f}}

\def\u{\hat{u}}
\def\u{\mathbf{u}}
\def\v{\mathbf{v}}

\def\x{\mathbf{y}}
\def\g{\mathbf{g}}

\def\transp{\text{T}}
\def\Ri{\text{Ri}}

\newcommand{\bit}{\vphantom{\dot{W}}}

\def\x{{\bf x}}

\begin{document}

\title{Optimal excitation of two dimensional Holmboe instabilities}

\author{Navid C. \surname{Constantinou}}
\email{navidcon@phys.uoa.gr}
\affiliation{
Department of Physics, University of Athens\\
Panepistimiopolis, Zografos 15784, Greece}
 \author{Petros J. \surname{Ioannou}}
\email{pjioannou@phys.uoa.gr}
\affiliation{
Department of Physics, University of Athens\\
Panepistimiopolis, Zografos 15784, Greece}

\begin{abstract}
\noindent(Under consideration for publication in Physics of Fluids.)\\
(Submitted on 27 Oct 2010 (v1), last revised 14 Apr 2011 (this version, v2))
\\

Highly stratified shear layers are rendered unstable  even at high stratifications by Holmboe instabilities when the density stratification is concentrated in a small region of the shear layer. These instabilities may cause mixing in highly stratified environments. However these instabilities  occur in limited bands in the parameter space. 
We  perform Generalized Stability  analysis of the two dimensional perturbation dynamics of an inviscid  Boussinesq stratified shear layer and show that Holmboe instabilities at high Richardson numbers can be excited by  their adjoints  at amplitudes that are orders of magnitude larger than by introducing initially the unstable mode itself. We also determine the optimal growth that is obtained for parameters for which there is no instability. We find  that there is potential for large transient growth  regardless of whether  the background flow is exponentially stable or not and that the  characteristic structure of the Holmboe instability asymptotically emerges as a persistent quasi-mode for parameter values for which the flow is stable.
\end{abstract}

\maketitle

\section{Introduction}

The mixing of shear layers and the development of turbulence  is severely impeded when the layer is located in regions of large stable stratification. The stratification is usually quantified with the non-dimensional  Richardson number  defined as the ratio of the local Brunt-V\"ais\"al\"a frequency $N^2$ to the square of the local shear $U'$, $\Ri = N^2/ U'^2$.
Large stratification corresponds to large Richardson numbers. The significance of the Richardson number for the stability of stratified flows has been underscored with a theorem due to Miles and Howard~\cite{Miles-1961,Howard-1961} which proves that if everywhere in the flow  $\Ri > 1/4$,  the flow is necessarily  stable to exponential instability.
The essential instability that leads primarily to mixing in both stratified and unstratified shear layers is the Kelvin-Helmholtz (KH) instability~\cite{Drazin-Reid-81}.
The KH modes are eventually stabilized when the local Richardson number exceeds $1/4$,  but if the stratification is concentrated in narrow regions within the shear layer the Richardson number may locally become smaller than $1/4$  and an instability can result although the overall stratification is very large. Under such conditions,  a new branch of instability of stratified shear layers  emerges as shown by Holmboe~\cite{Holmboe-1962}, consisting of  a pair of propagating waves with respect to the center flow, one prograde and one retrograde. This new  instability branch has been named the Holmboe (H) instability and it is physically very interesting because it persists at all stratifications and can lead to mixing in highly stratified shear layers.   

Holmboe instabilities have been reproduced in laboratory experiments~\cite{Thorpe-1971,Browand-Winant-1973,Pouliquen-etal-1994,Caulfield-etal-1995,Zhu-Lawrence-2001,Hogg-Ivey-2003, Tedford-etal-2009} and have been numerically simulated~\cite{Smyth-etal-1988,Sutherland-etal-1994,Smyth-2006,Smyth-Peltier-2006,Smyth-etal-2007,Carpenter-etal-2007,Alexakis-2009,Tedford-2009}. At finite amplitude these Holmboe modes equilibrate into propagating waves and  they can induce mixing in highly stratified environments~\cite{Smyth-etal-1988,Smyth-Winters-2003,Carpenter-etal-2007}. Consequently, Holmboe instabilities present  the intriguing possibility that they may be  responsible for transition to turbulence and mixing in highly stratified flows.  In that vein, Alexakis~\cite{Alexakis-2005,Alexakis-2007,Alexakis-2009} has proposed that  the surmised necessary  mixing in order for a thermonuclear runway to occur  and  highly stratified white dwarfs  form supernova explosions could be effected with Holmboe instabilities.

Holmboe~\cite{Holmboe-1962} also  introduced a new method for analyzing the perplexing instabilities that may arise with the introduction of stratification. He went beyond classical modal analysis and proposed that a fruitful program for predicting and obtaining a clear physical idea of the instability of stratified shear flows is to simplify the background flow to  segments of piecewise constant vorticity, as in  Rayleigh's seminal work~\cite{Rayleigh-1880}, and then consider the dynamics of the edge waves that are supported at each vorticity and density discontinuity. Holmboe showed that the flow becomes unstable  when the edge waves propagate with the same phase speed.
This method of analysis has been extended and used to clarify the detailed mechanism of instability in stratified flows~\cite{Baines-Mitsudera-1994,Caulfield-1994,Haigh-Lawrence-1999,Harnik-etal-2008,Rabinovich-etal-2011} and has been shown recently by  Carpenter et al.~\cite{Carpenter-etal-2010} to be capable of readily assessing the character of instability of non-idealized observed flows. 
Also this edge wave description has  elucidated the dynamics of other instabilities that occur in geophysics~\cite{Bretherton-1966,Sakai-1989,Heifetz-Methven-2005, Bakas-Ioannou-2009} and  astrophysics ~\cite{Goldreich-etal-1986,Narayan-etal-1987,Umurhan-2010}.

However, the efficacy  of Holmboe modes to mix highly stratified layers  can be questioned  because the instability occurs only in limited regions of parameter space and  the modal growth rate of the instability becomes  exponentially small with stratification.
Specific estimates of the asymptotic behavior of the growth rates and of the unstable band of wavenumbers has been recently obtained by Alexakis~\cite{Alexakis-2007}.
Also the introduction of viscosity can further reduce the growth rates of the  Holmboe instabilities. 
It is the purpose of this work to go beyond the modal analysis and investigate the non-modal  stability of stratified shear layers that are Holmboe unstable in order to assess the true potential of growth of stratified shear layers. We will achieve this using the standard methods of  generalized stability theory~\cite{Farrell-Ioannou-1996a,Schmid-Henningson-2001}.

Because shear layers are powerful transient amplifiers of perturbation energy, we expect  that the Holmboe instabilities can be excited at enhanced  amplitude by composite  non-modal perturbations. 
Even in the case of an infinite  constant shear flow which has a continuous spectrum with no inviscid analytic modes  and all perturbations eventually decay algebraically with time, the non-modal solutions constructed by Kelvin~\cite{Kelvin-1887b} demonstrate that the asymptotic limit is non-uniform and perturbation energy can transiently exceed any chosen bound in the inviscid limit; the same is true for bounded Couette flow as shown by Orr~\cite{Orr-1907}. The Kelvin-Orr solutions can be extended to stratified flows~\cite{Phillips-1966,Hartmann-1975} and these solutions can produce  transient perturbation growth that  can also exceed in the inviscid limit any bound~\cite{Farrell-Ioannou-1993c}. Specifically it can be shown that 
in constant shear flow  and large Richardson numbers  the  perturbation energy amplification  
is the square root of that achieved in the absence of stratification. 
The same robust growth resulting from the continuous spectrum is expected to persist in  all shear layers as the dynamics of the Orr waves are universal and do not depend on the details of the background shear flow~\cite{Farrell-Ioannou-1993b}.

Furthermore in shear layers of finite size the vorticity and density edge waves that are concentrated in  regions of  vorticity and density discontinuities can interact  and produce  transient perturbation growth~\cite{Heifetz-etal-1999}. 
Also the transiently growing Kelvin-Orr vorticity structures may deposit their energy to the edge waves~\cite{Bakas-Ioannou-2009}  or to propagating  gravity waves~\cite{Lott-1997,Bakas-Ioannou-2007,Lott-etal-2010}.
In general the continuous spectrum can excite at high amplitude the unstable or even stable  analytic modes~\cite{Farrell-1988b,Farrell-1989} and maintain under continuous excitation high levels of variance in stable flows~\cite{Farrell-Ioannou-1993d,Farrell-Ioannou-1994b}.

In this work we will consider the generalized stability of a Boussinesq stratified shear layer. The Boussinesq approximation makes inaccurate predictions of the dispersion relation of the perturbations in the shear layers when the stratification is locally very large~\cite{Umurhan-Heifetz-2007}.  However  we have found that the resulting optimal growths are not sensitive to this approximation and in this paper  in order to make contact  with previous work we  will only present results obtained using the Boussinesq approximation. 
Also following Hazel~\cite{Hazel-1972} and the work of Smyth and Peltier~\cite{Smyth-Peltier-1989,Smyth-Peltier-1990} we will consider a continuous version of  Holmboe's  velocity profile with the  characteristic density stratification embedded in  the shear layer. Hazel and Smyth and Peltier had analyzed the character of the bifurcations that occur in the instability of the continuous profile as the Richardson number at the center of the shear layer increases. Alexakis~\cite{Alexakis-2005,Alexakis-2007,Alexakis-2009} extended the analysis to even higher stratifications and was able to predict theoretically and show numerically that there are higher branches of Holmboe instabilities in these continuous profiles. 
Because of the geophysical and astrophysical interest we will concentrate our analysis of non-modal perturbation growth of shear layers at very high Richardson numbers which possess sparse islands of exponential instability of low growth rate.

\section{\label{sec:formulation} Formulation}

We will study the stability of  an inviscid, unidirectional, stratified,  step like  channel flow to 
two  dimensional incompressible perturbations. 
The mean flow is characterized by a velocity jump $\Delta U$ over a  vertical scale  $h_0$ and
the associated statically stable  mean density is characterized by a density jump $\Delta \rho$.
The linearized nondimensional perturbation equations 
about the mean flow $U_0(z)$ and mean density $\rho_0(z)$ in the Boussinesq approximation are:
\begin{subequations}
\label{eq:RHO_system_nondim}
\begin{align}
\(\partial_t + U_0\partial_x \) u &=-U_0'\,w-\partial_x p\label{eq:RHO_u_nondim}\\
\(\partial_t + U_0\partial_x \) w &=-J\r-\partial_z p\label{eq:RHO_w_nondim}\\
\(\partial_t + U_0\partial_x\)\r &=-\rho_0'\, w  \label{eq:RHO_rho_nondim}\\
\partial_x u + \partial_z w &= 0 \label{eq:RHO_cont_nondim}
\end{align}
\end{subequations}
The density of the fluid has been decomposed as: $\rho\equiv\rho_m + \rho_0(z) + \r(x,z,t)$, where $\rho_m$ is the mean background density, $\rho_0(z)$ is the mean density variation and $\r(x,z,t)$ is the perturbation density; and furthermore, according to the Boussinesq approximation, we assume, $|\rho_0|\ll \rho_m$ and $|\r|\ll\rho_m$.
We denote with  $u$ and  $w$  the perturbation velocities in the streamwise ($x$) and vertical ($z$) direction and  $p$ is the perturbation reduced pressure. 
Differentiation with respect to $z$ is denoted with a dash. The perturbation equations have been made nondimensional, choosing $h_0$ as  the length scale, $\Delta U$  as the velocity scale, and $\Delta \rho$ as the scale for the density. Time has been scaled with  $h_0/\D U$, and pressure has been scaled with $\rho_m\(\D U\)^2$. The local Richardson number is
\be
\Ri = - J\frac{\rho_0'}{U_0'^2}
\label{eq:ri}
\ee
with
\be
J=\frac{g\,\D\rho\,h_0}{\rho_m\,\D U^2}~.
\label{eq:def_J}
\ee
$J$ provides a measure of the bulk Richardson number of the shear layer. We will assume that the flow is in a channel of length $L=2h_0$ and impose at the channel boundaries zero vertical velocity, $w$, and zero  perturbation density, $\r$. The length of the channel has been selected so that boundaries do not affect in an important way the results that we report in this
paper. We have checked by doubling the channel size that the  eigenspectrum 
and the perturbation transient growth 
are converged. Similar insensitivity to the location of the channel walls has been previously  reported by Alexakis~\cite{Alexakis-2009}.

The perturbation equations can be written compactly by introducing a streamfunction $\psi$ so that the velocities are $u=\partial_z\psi$ and $w=-\partial_x\psi$, and the vertical displacement, defined through the relation:
\be
\(\partial_t + U_0\partial_x \)\eta = w~,\label{eq:def:eta_nondim}
\ee
can replace the perturbation density  as:
\be
\r = -\rho_0'\,\eta~.
\label{eq:def_eta}
\ee
The perturbation state will be  determined by the fields $\psi$ and $\eta$.
Because of homogeneity in the streamwise direction,  the perturbation state can be decomposed in Fourier components in $x$: 
\be
\hat{\x}(z,t) e^{i kx}= [\hat{\psi}(z,t),\hat{\eta}(z,t)]^\transp e^{i kx}
\ee
and the perturbation dynamics for the Fourier amplitudes for each wavenumber $k$ can be written compactly in the form:
\be
\partialf{\,\hat{\x}}{t}  = \A\,\hat{\x}~,
\label{eq:system_nondim_modal}
\ee
with the operator $\A$ given by:
\be
\label{eq:def_operator_A}
\A = -ik \[
\begin{array}{cc}
 \D^{-1}\(  U_0\D - U_0'' \) ~~~~~  &    J\D^{-1}\rho_0'\\
  1  &    U_0
\end{array}
\]~.
\ee
With   $\D =  \partial_{zz}^2-k^2$ we denote the two dimensional Laplacian for the $k$ streamwise wavenumber.
$\D^{-1}$  is the inverse Laplacian, the inversion being rendered possible and unique by imposing the boundary conditions on the upper and lower boundaries. 
The perturbation equation \eqref{eq:system_nondim_modal} is equivalent to the time dependent Taylor-Goldstein equation:
\be
\label{eq:Taylor_Goldstein}
\(\partial_t + ik U_0 \)^2\D \hat{\psi}  - ik U_0''\(\partial_t + ik U_0 \) \hat{\psi} +k^2 J\rho_0' \hat{\psi}=0~.
\ee

\section{Kelvin-Helmholtz and Holmboe instabilities of a shear layer}

\begin{figure*}
\includegraphics{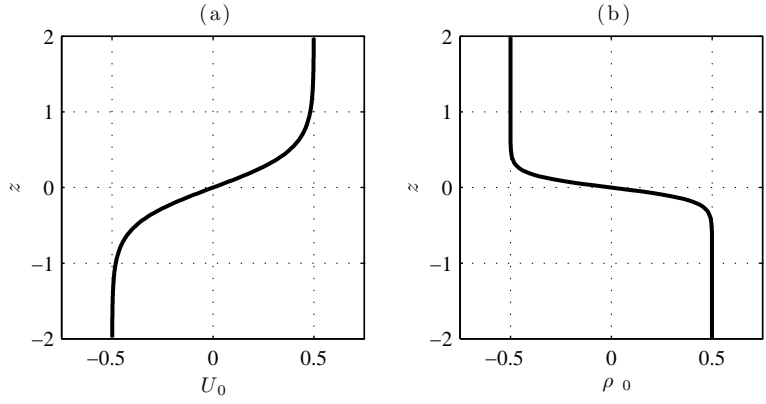}
\caption{\label{fig:main_profiles}(a): Mean velocity profile as a function of height. (b): Mean density profile as a function of height for
$R=3$.}
\end{figure*}

Following Hazel~\cite{Hazel-1972}, we consider the following background mean velocity and density:
\be
\label{eq:main_profiles_holmboe}
U_0(z) = \frac{1}{2}\tanh{\(2z\)}  ~~,~~ \rho_0(z) =- \frac1{2}\tanh{\(2Rz\)}~.
\ee
The mean velocity and density profiles are shown in FIG.~\ref{fig:main_profiles} for $R=3$. The parameter $R$ is the ratio of the width of the shear layer to the width of the region over which the density jump occurs.
The local Richardson number is given by,
\be
\Ri(z)=R J \frac{\sech^2{\(2Rz\)}}{\sech^4{\(2z\)}}~.
\ee
The Richardson number at the center is related to the bulk Richardson number through the relation:
\be
\Ri(0) = R J~. 
\ee

\begin{figure*}
\includegraphics{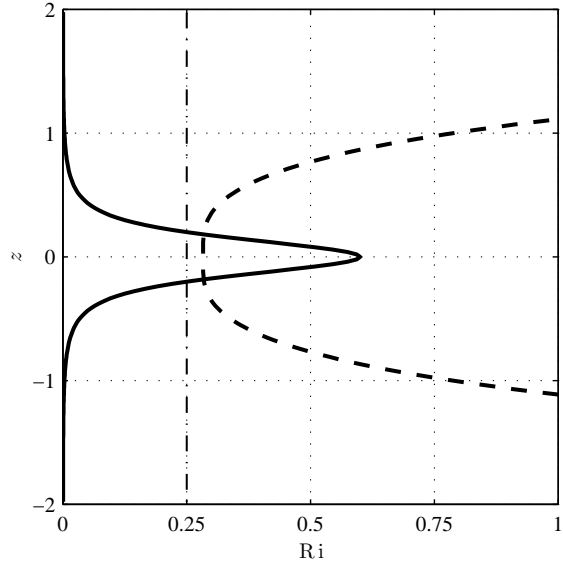}
\caption{\label{fig:Richardson} Richardson number as a function of height for the velocity and density profiles in FIG.~\ref{fig:main_profiles} for $J=0.2$ and $R=3$ (solid). The Richardson number assumes values smaller than $1/4$ away from the center and the flow may be exponentially unstable. Also shown is the case of $J=0.2$, $R=\sqrt{2}$ (dashed). In this case the Richardson number is everywhere greater than $1/4$ and the velocity profile is by necessity exponentially stable. The threshold value $\Ri=1/4$ (dash-dot) is also  indicated.}
\end{figure*}

When $R=1$ the local Richardson number assumes its minimum value at the center, $\Ri(0)=RJ$, and the shear layer is unstable only if $\Ri(0)<1/4$. The instability that results  is a Kelvin-Helmholtz (KH) type instability with a short wave cutoff and zero phase velocity. For $R>2$ the Richardson number decays exponentially to zero. This exponential decay allows the local Richardson number to be in regions smaller than $1/4$ and the shear layer may become unstable although both the bulk stratification $J$  and $\Ri(0)$  may be very large. The instabilities that occur under these conditions are the Holmboe instabilities (H). The variation of the Richardson number with height is shown in  FIG.~\ref{fig:Richardson} for $R=3$, a case that supports Holmboe instabilities and will be treated in detail in  this paper, and for $R=\sqrt{2}$ which can support only KH instability when the Richardson number  at the center is smaller than $1/4$.

In order to determine the modal stability of the velocity and density profiles \eqref{eq:main_profiles_holmboe} we assume modal solutions of \eqref{eq:system_nondim_modal} of the form $\hat{\x}(z,t)= [\hat{\psi}_c(z),\hat{\eta}_c(z)]^\transp e^{-i kc t}$, with  $-i k c$  an eigenvalue of the evolution operator $\A$ in (\ref{eq:def_operator_A}). The modal growth rate of the perturbation is $k c_i$ where $c_i=\Im (c)$ and the flow is exponentially unstable to perturbations of wavenumber $k$ if $c_i >0$.  The real part, $c_r$,  of the eigenvalue $c(k)$, gives the phase speed of the perturbation.
Detailed analysis of the eigenspectrum of \eqref{eq:main_profiles_holmboe}  and the bifurcation from KH instabilities to H instabilities has been already performed~\cite{Hazel-1972,Smyth-Peltier-1989, Smyth-Peltier-1990,Alexakis-2005,Alexakis-2007, Alexakis-2009}. Here we review the basic stability characteristics for the case of $R=3$ that admits both KH and H instability modes.

\begin{figure*}
\includegraphics{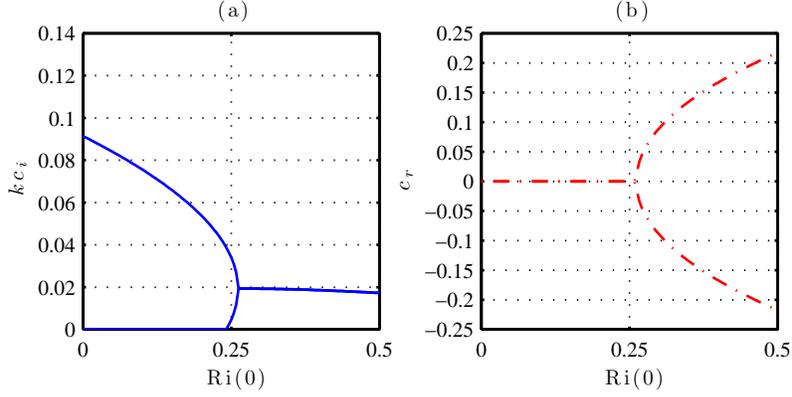}
\caption{\label{fig:growth_rates_cr_N_401_R_3_k_0p3_Ri0max_0p5} (Color online) Bifurcation properties of the unstable modes as a function of the Richardson number at the center of the shear zone $\Ri(0)=R J$  for the mean state (\ref{eq:main_profiles_holmboe}) for $R=3$ and  perturbations with $k=0.3$.
 (a) the growth rate $k c_i$ (solid); (b) the associated phase speed $c_r$ (dashed-dot). For $\Ri(0)< 0.242 $ there is a single Kelvin-Helmholtz (KH) mode with zero phase speed. For $0.242<\Ri(0)<0.258$ there are two KH modes with zero phase speed. The two KH modes coalesce to produce two unstable Holmboe (H) modes with equal and opposite phase speeds.}
\end{figure*}

The transition from KH instability to H instability is shown  in  FIG.~\ref{fig:growth_rates_cr_N_401_R_3_k_0p3_Ri0max_0p5}  where we plot the 
growth rate and phase speed as a function of the Richardson number at the center of the shear zone  $\Ri(0)$ for $k=0.3$.
For   $\Ri(0) < 0.242$, there is a single unstable Kelvin-Helmholtz instability with phase speed   $c_r=0$. At $\Ri(0)=0.242$ a second KH mode becomes unstable, also with $c_r=0$.
The KH modes coalesce at $\Ri(0)=0.258$ to form a propagating pair of unstable Holmboe modes with equal and opposite phase speeds and the same growth rate. 
In this case we see  two branches of KH instabilities existing even when the center Richardson number exceeds $1/4$ (the Richardson number assumes values smaller than $1/4$ way from the center because the parameter is $R=3$, see FIG.~\ref{fig:Richardson});  a similar case can be found in Smyth and Peltier~\cite{Smyth-Peltier-1989}.

Numerical results were obtained by discretizing the channel and all differential operators 
using second order centered differences and incorporating the appropriate boundary conditions.  
Equation  \eqref{eq:system_nondim_modal}  then becomes a
matrix equation in which the state becomes a column vector. 
We used a staggered grid, that is, we evaluated $\hat{\psi}_c(z)$ at $N$ interior points in the vertical
and $\hat{\eta}_c(z)$ at $N+1$ points located  halfway between the collocation points of the streamfunction. 
The eigenspectrum is  obtained by eigenanalysis of  operator $\A$. 
Calculations, unless otherwise specified, were performed with $N=1001$ in the domain $-2 \le z \le 2$.
With this discretization the finite time evolution of the dynamics is well resolved  when there is no instability
up to a time of $t=O(1/(\alpha k \Delta z))$ where $\alpha$ is the typical shear\cite{Farrell-Ioannou-1993c}. However, in order to  numerically resolve
the curves  of  zero  modal growth  of the Holmboe modes of instability 
we had to include  numerical diffusion in both the momentum
and density equations  with coefficient of diffusion $\nu = (\Delta z/2)^2$ in the momentum equation and $\nu/9$ in the 
density equation for the case $R=3$ ($\Delta z$ is the grid spacing).
Very accurate determination of the neutral curve in the inviscid limit can be obtained, 
using  the shooting method as in  Alexakis~\cite{Alexakis-2007}.

\begin{figure*}
\includegraphics{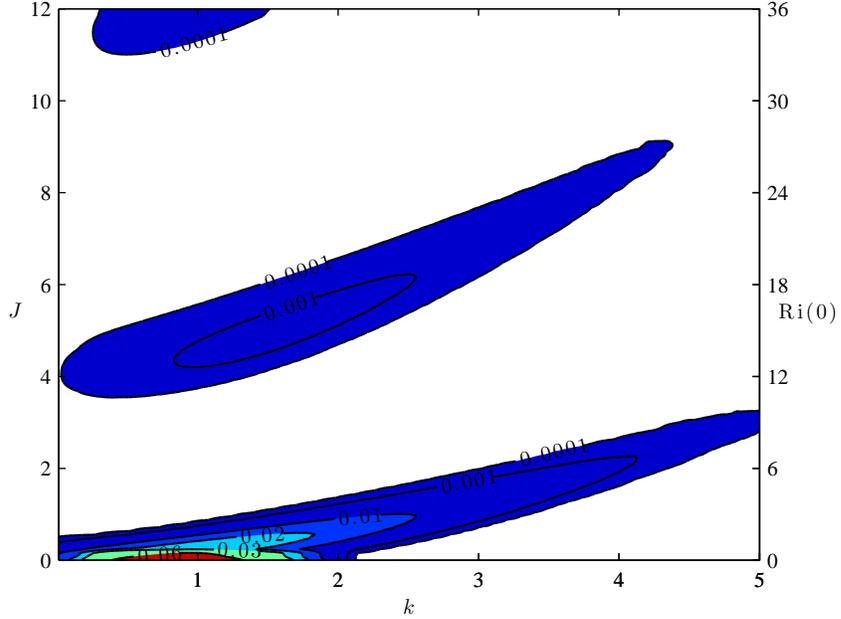}
\caption{\label{fig:spectrum_R_3_LARGE}(Color online) Contours of modal growth rate $kc_i$ of the perturbation operator $\A$ as a function of wavenumber, $k$, and the bulk Richardson number, $J$, for $R=3$. Because of the inclusion of diffusion the instability bands
do not extend to infinity as the do in the inviscid limit.
The corresponding value of the Richardson number at the origin $\Ri(0)=R J $ is also recorded  at the ordinate axis on the right. For this value of $R$ there are  KH unstable modes with phase speed $c_r=0$ for small values of $J$ (barely discernible in this plot) and three unstable bands of pairs of H instabilities with prograde and retrograde phase speeds for higher $J$.}
\end{figure*}

Growth rate contours as a function of stratification,  $J$ or $\Ri(0)$,  and wavenumber, $k$
are shown in FIG.~\ref{fig:spectrum_R_3_LARGE}
for $R=3$. 
The region of KH instability is concentrated for small Richardson numbers and it is not 
discernible in this plot. However the bands of instability  corresponding to the various Holmboe 
modes of
instability can be clearly seen.  
Note that because of the inclusion of diffusion the instability bands do not extend to infinity
as they do in the inviscid limit~\cite{Holmboe-1962,Howard-Maslowe-1973,Baines-Mitsudera-1994,Alexakis-2005}. 

\begin{figure*}
\includegraphics{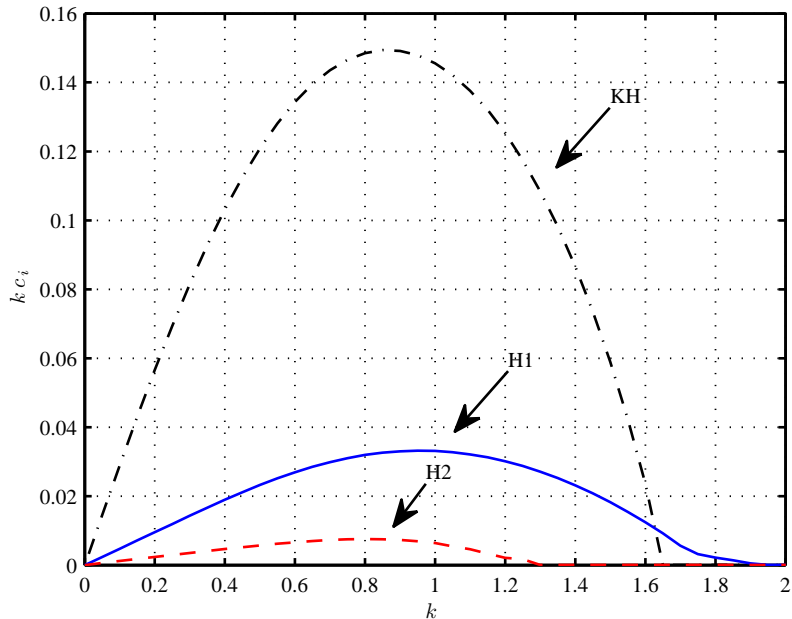}
\caption{\label{fig:spectrum_plot_four_states}(Color online) Growth rates $kc_i$ as a function of wavenumber $k$ for: a  KH instability for $\Ri(0)=0.06$ (dash-dot), an  H1 mode of instability for $\Ri(0)=0.65$ (solid) and an  H2  mode of instability for $\Ri(0)=12$ (dashed). The growth rates of H2 mode  have been multiplied by 10.}
\end{figure*}

The first band in FIG.~\ref{fig:spectrum_R_3_LARGE} 
 is the first Holmboe mode of instability, H1, (for $J>0.0833$), the second is the second Holmboe instability mode, H2, (for $J>3.5$) and we can also  partly see  the growth rate of  the third Holmboe instability mode, H3 (for $J>10$). Estimates of the growth rates of these higher modes and regions of instability can be found in Alexakis~\cite{Alexakis-2007} who shows that the growth rate of the Holmboe instabilities 
is decreasing exponentially with stratification and  as the 
Holmboe modes approach neutrality  the  critical layer
of the instabilities, that is the height $z_c$ for which $U_0(z_c)=c_r$, tends  to the height that corresponds to the maximum/minimum velocity  $\pm 0.5$ of the shear zone~\cite{Alexakis-2007}. 
We obtain a sense of the ordering of the growth rates of the KH and H instabilities in FIG.~\ref{fig:spectrum_plot_four_states}. In that figure  we plot the modal growth rates $kc_i$ as a function of wavenumber $k$ for $R=3$ and  center  Richardson numbers  $\Ri(0)=0.06,\ 0.65,\ 12$ for which we have respectively KH mode instability,  H1  and H2 instabilities.

\begin{figure*}
\includegraphics{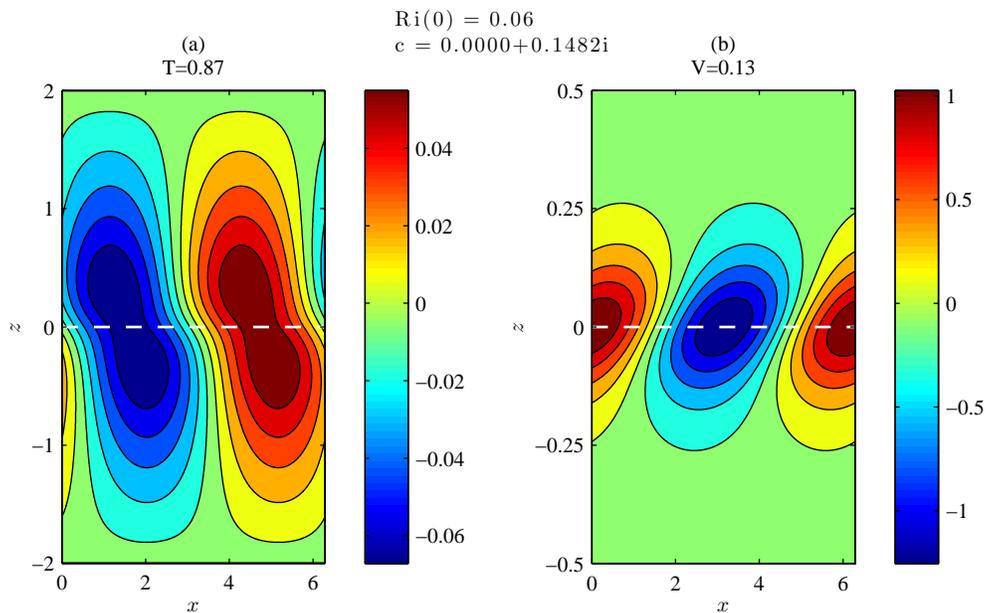}
\caption{\label{fig:KH_mode_Ri0_0p06}(Color online) Structure of the streamfunction (panel (a)) and density   (panel (b)) perturbation fields   of an unstable KH  mode with growth rate  $kc_i = 0.1482$ and phase speed $c_r=0$. Critical layer at $z_c=0$ (dashed). For perturbations with $k=1$, and base flow with $R=3$ and $\Ri(0)=0.06$.}
\end{figure*}

\begin{figure*}
\includegraphics{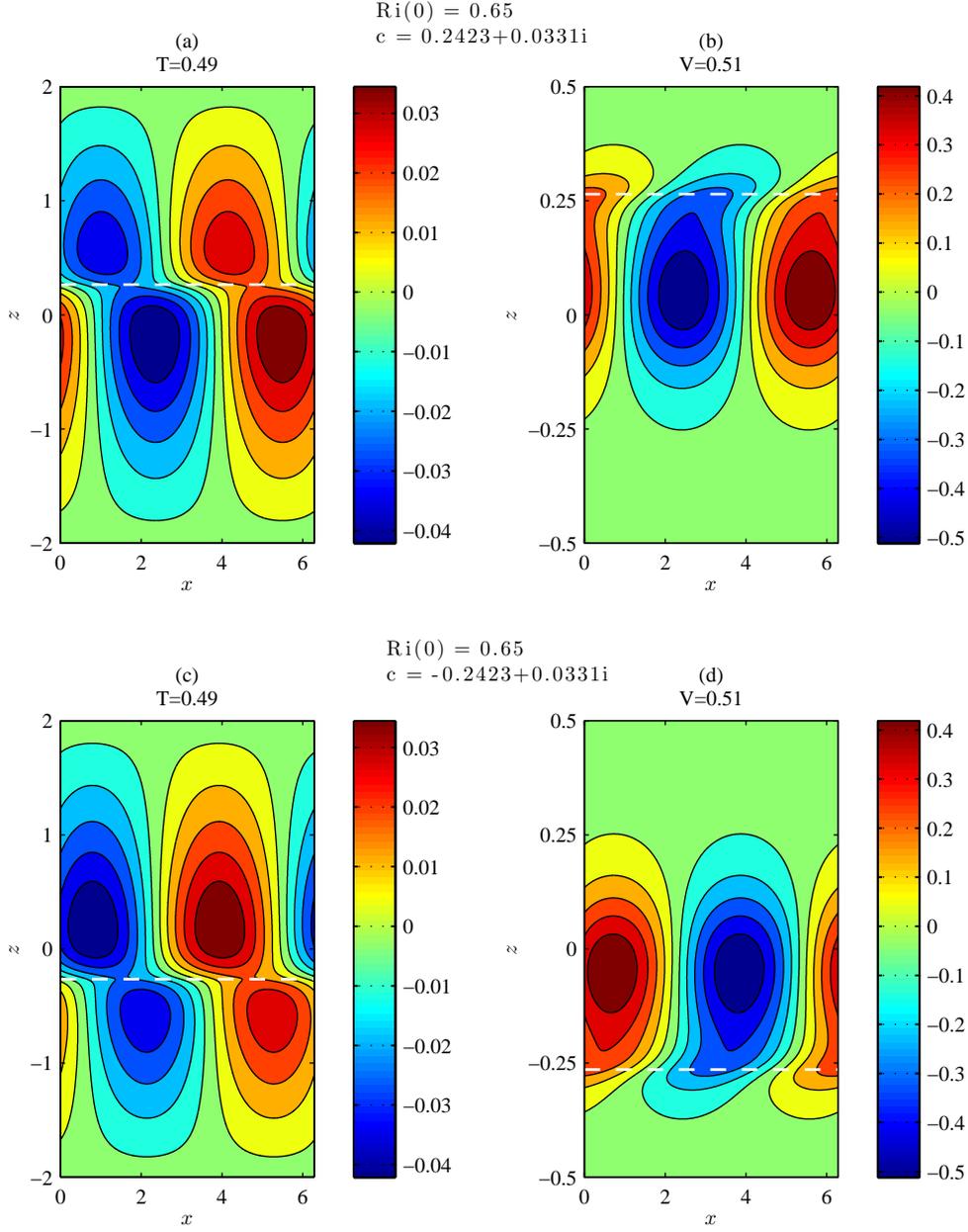}
\caption{\label{fig:Holmboe_mode1}(Color online) Structure of the streamfunction and density perturbation fields of  unstable H1  modes with growth rate  $kc_i = 0.0331$ and 
phase speeds $c_r=\pm0.2423$. Panels (a) and (b):  streamfunction and density contours  for the  prograde H1 mode. Panels (c) and (d): for the retrograde H1 mode. Critical layer at $z_c=\pm 0.264$ (dashed). For perturbations with $k=1$, and base flow with $R=3$ and $\Ri(0)=0.65$.}
\end{figure*}

Typical structure of the KH instability mode is shown in FIG.~\ref{fig:KH_mode_Ri0_0p06} for $k=1$, $R=3$ and  $\Ri(0)=0.06$. For these parameters the coalescence of the KH modes and the emergence of the first Holmboe mode of instability occurs at $\Ri(0)=0.182$. The Holmboe modes come in prograde and retrograde pairs. 
The prograde and retrograde unstable Holmboe modes, H1, for $\Ri(0)=0.65$ are shown in  FIG.~\ref{fig:Holmboe_mode1}. Because they are symmetric with respect to $z=0$, we  will subsequently plot only the prograde branch of the H mode.
Higher Holmboe modes have larger vertical  wavenumbers, the case of H2  is shown in FIG.~\ref{fig:Holmboe_mode2}.  In the  idealized piecewise constant problem studied by Holmboe (see Appendix~\ref{sec:Holmboe_classical}) the prograde Holmboe instability arises from  resonant interaction between the edge waves at the density discontinuity at the center with the edge wave at the  upper vorticity
discontinuity, and  with the edge wave at the lower vorticity  discontinuity for the retrograde 
instability\cite{Baines-Mitsudera-1994}. 
The same type of interaction gives rise to the Holmboe instability modes when the discontinuities  of the background  are smoothed as for the background given by Eq.~(\ref{eq:main_profiles_holmboe}). Because of the  delta function nature of the vorticity and stratification
gradient the edge waves in  Holmboe's idealized profile do not have internal structure
and as a result there is a single  pair of unstable Holmboe instabilities that can arise from the
interaction. The smoothed profile allows the edge wave to obtain vertical structure 
within the regions of large vorticity and stratification gradient (cf. respectively to the left and right panels of  FIG.~\ref{fig:Holmboe_mode1} and FIG.~\ref{fig:Holmboe_mode2})  
and the H2, H3, etc  Holmboe  instabilities emerge, with the  higher modes associated with higher  vertical wavenumber structure.

\section{Optimal excitation of Kelvin-Helmholtz and Holmboe modes}

The perturbation dynamics in a stratified shear flow are non-normal and, as we will show in this section,  the unstable modes emerge through excitation of their adjoint. In order to determine the
growth that results by introducing initially the adjoint of the unstable mode
we derive  the adjoint perturbation dynamics in the energy norm. 
For a  perturbation of wavenumber $k$ the total average kinetic energy  over a wavelength in the region $[z_1,z_2]$ is:
\be
T = \frac1{4} \int\limits_{z_1}^{z_2}\!\df z\; \hat{\psi}^*\(-\D\hat{\psi}\)~,
\ee
and the potential energy is:
\be
V = \frac1{4} \int\limits_{z_1}^{z_2}\!\df z\;J \(-\rho_0'\)\hat{\eta}^*\hat{\eta} ~,
\ee
The total energy of a perturbation  $\hat{\x}=[\hat{\psi}, \hat{\eta}]^\transp$ is thus given by the inner product:
\be
|| \hat{\x} ||^2 =(\hat{\x},\hat{\x})= \frac1{4} \int_{z_1}^{z_2} \df z \;\[\hat{\psi}^*\,\(-\D\hat{\psi}\) +  J \(-\rho_0'\)\hat{\eta}^*\hat{\eta} \]~.
\label{eq:def_norm}
\ee 
The adjoint operator $\A^{\dagger}$ in this inner product is defined as the operator that satisfies:
\begin{equation}
\label{eq:def_adjoint}
\( \bit \A^\dag \hat{\fb}_\a, \hat{\g} \)  = \(  \bit \hat{\fb}_\a, \A\,\hat{\g} \)   ~,
\end{equation}
for any two states $\hat{\fb}_\a=[\hat{\psi}_\a,\hat{\eta}_\a]^\transp$ and $\hat{\g}=[\hat{\psi},\hat{\eta}]^\transp$. 
For a discussion of the adjoint equation in fluid mechanics  see Farrell~\cite{Farrell-1988a}, Hill~\cite{Hill-95}, 
Farrell and Ioannou~\cite{Farrell-Ioannou-1996a}, Chomaz et. al.~\cite{Chomaz-etal-2009}  and
Schmid and Henningson~\cite{Schmid-Henningson-2001}.

From (\ref{eq:def_adjoint})  the adjoint operator of (\ref{eq:def_operator_A}) in the energy inner product  is easily shown to be:
\be
\label{eq:def_operator_Aadj}
\A^\dag = ik \[
\begin{array}{cc}
 \D^{-1}\(  U_0\D + 2U_0'\,\Df \) ~~~~~  &    J\D^{-1}\rho_0'\\
  1  &    U_0
\end{array}
\]~,
\ee
with $\Df=\df/\df z$ denoting differentiation with respect to $z$, and the adjoint  evolution equation is 
\be
\partialf{\hat{\x}_\a}{t} = \A^\dag\,\hat{\x}_\a
\label{eq:ad_system_nondim_modal}
\ee
with the adjoint boundary conditions $\hat{\x}_{\a}(z_i,t)=0$ at the channel walls $z_i$. 
This equation produces the adjoint of the time-dependent Taylor-Goldstein equation \ (\ref{eq:Taylor_Goldstein}):
\be
\label{eq:ad_Taylor_Goldstein}
\(\partial_t - ik U_0 \)^2 \D \hat{\psi}_\a -2ikU_0'\(\partial_t - ik U_0 \)\Df \hat{\psi}_\a-k^2 J\rho_0' \hat{\psi}_\a=0~.
\ee
The eigenvalues of the adjoint operator  $\A^\dag$  are the complex conjugate of the eigenvalues  of $\A$ and in this way we can establish a correspondence between the modes of the adjoint $\A^\dag$ 
and the modes of $\A$. Further, the mode of the adjoint 
with eigenvalue $ikc^*$, $\hat{\x}_{\a\,c}$, and the mode of $\A$ with eigenvalue $-i k c'$, $\hat{\x}_{c'}$,  satisfy the biorthogonality relation:
\be
\(\bit\hat{\x}_{\a\,c},\hat{\x}_{c'}\) = 0\quad \text{for}\quad c\ne c'~.
\label{eq:biorth}
\ee
From (\ref{eq:ad_Taylor_Goldstein}) it can be  verified that for any analytic mode of the Taylor-Goldstein equation  \ (\ref{eq:Taylor_Goldstein}) with eigenvalue $-ikc$  (with $c_i>0$ or with $c_r$ outside the range of the background flow) 
the  mode of  $\A^\dag$ with eigenvalue $ikc^*$ is:
\be
\hat{\x}_{\a\,c} = \[
\begin{array}{c}
\dfrac{\hat{\psi}_{c}^*}{U_0-c^*} \\
\dfrac{\hat{\eta}_{c}^*}{U_0-c^*}
\end{array}
\]~.
\label{eq:adjoint_analytic_expressions}
\ee

From the biorthogonality relation (\ref{eq:biorth}) arises the physical importance of the modes of the adjoint: of all perturbations 
of unit energy  the largest projection on a given mode is achieved by the adjoint of the mode.
Indeed,  a state written as a superposition  of the modes of $\A$:
\be
\hat{\x}= \sum_c a_c \,\hat{\x}_c~,
\label{eq:decomposition_of_general_state}
\ee
will have coefficients  $a_c$ given by:
\be
a_c = \frac{\(\hat{\x}_{\a\,c},\hat{\x}\)}{(\hat{\x}_{\a\,c},\hat{\x}_c)}~.
\label{eq:decomposition_coeff_of_general_state}
\ee
Because $|\(\hat{\x}_{\a\,c},\hat{\x}\)| \le 1 $, if the mode and the state $\hat \x$ are normalized, with equality when $\hat{\x}$ and $\hat{\x}_{\a\,c}$ are multiple to each other, 
the coefficients satisfy the  inequality:
\be
|a_c| \le \frac{1}{|(\hat{\x}_{\a\,c},\hat{\x}_c)|}~,
\label{eq:decomposition_coeff_of_general_state}
\ee
and the maximum  projection on a given mode is achieved  by  the adjoint mode, i.e. by choosing $\hat{\x}=\hat{\x}_{\a\,c}$. 
The adjoint perturbation excites the mode at an energy which is a factor 
\be
\frac{1}{|(\hat{\x}_{\a\,c},\hat{\x}_c)|^2}~,
\label{eq:amplification}
\ee
greater than an initial condition consisting of the mode itself. For highly non-normal systems  $|(\hat{\x}_{\a\,c},\hat{\x}_c)|\ll 1$ and as a result this amplification may be very large, implying that the emergence of the mode in these systems is mainly due to the excitation of the adjoint. Although this is not a new result, it has not been  noted in previous studies concerning Holmboe instabilities.

\begin{figure*}
\includegraphics{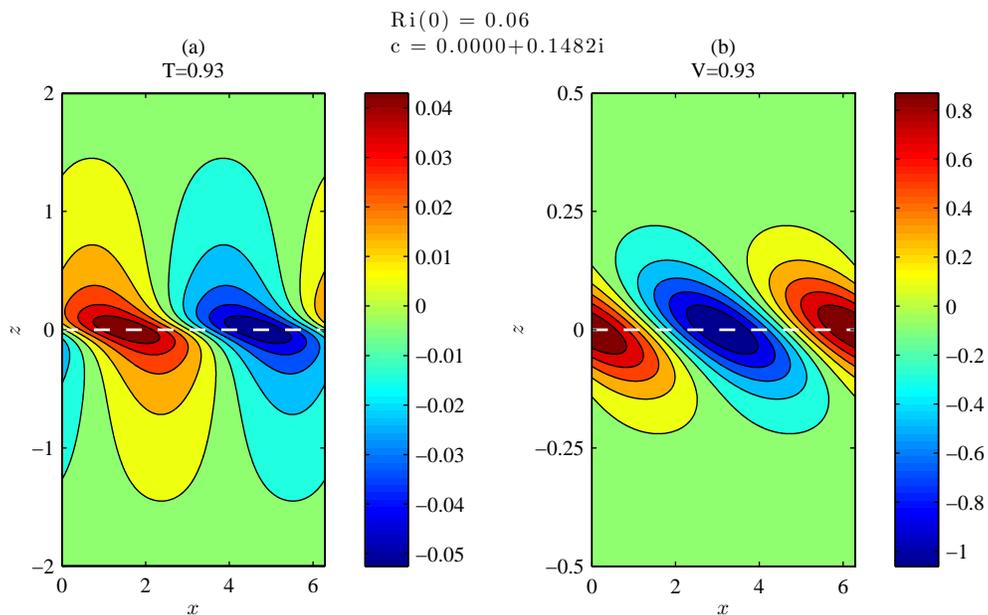}
\caption{\label{fig:KH_mode_Ri0_0p06_adjoint}(Color online) Structure of the streamfunction (panel (a))  and density  (panel (b)) perturbation fields of the adjoint of the unstable KH mode of FIG.~\ref{fig:KH_mode_Ri0_0p06}. Parameters as in FIG.~\ref{fig:KH_mode_Ri0_0p06}.}
\end{figure*}

\begin{figure*}
\includegraphics{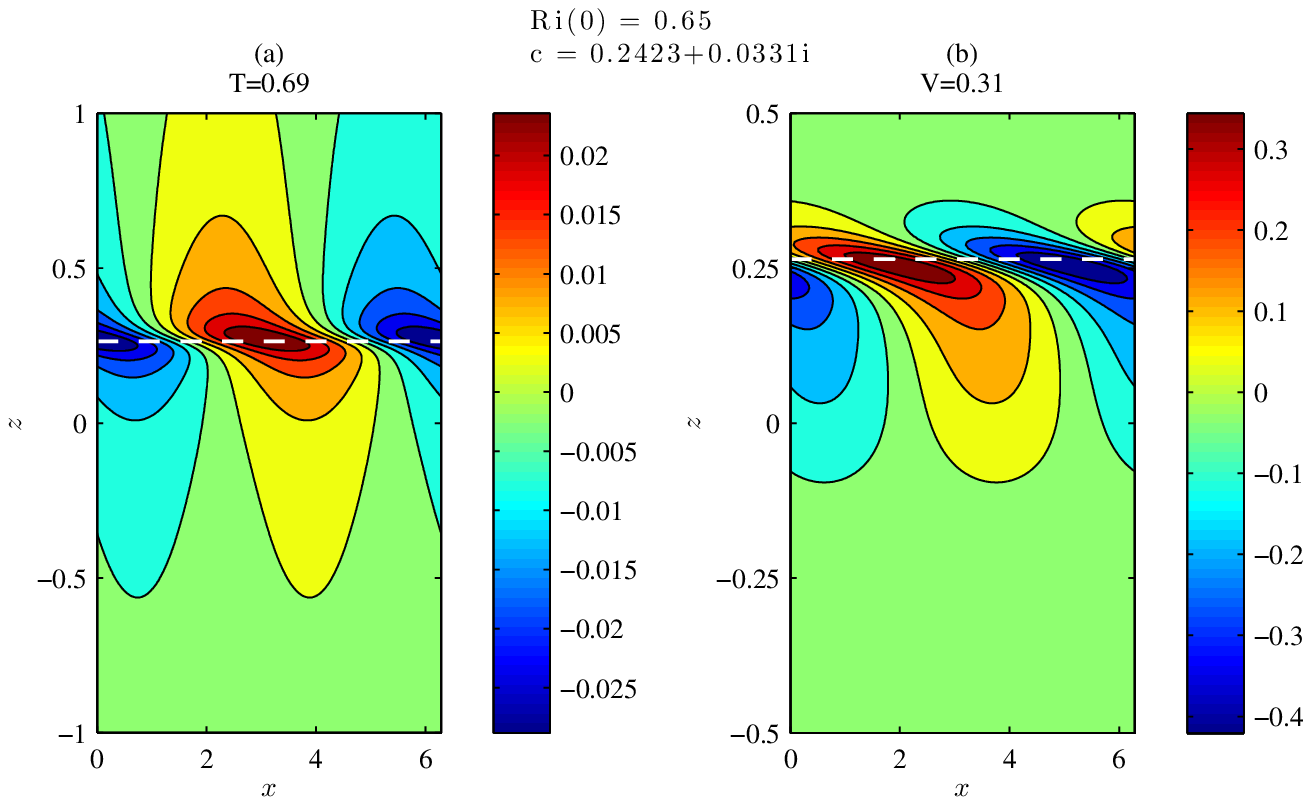}
\caption{\label{fig:Holmboe_mode1_adjoints}(Color online) Structure of the streamfunction and density perturbation fields of the adjoint of the unstable prograde H1 mode of FIG.~\ref{fig:Holmboe_mode1}. The adjoint is centered at the steering level, $z_c=0.264$,  of the mode.}
\end{figure*}

In FIG.~\ref{fig:KH_mode_Ri0_0p06_adjoint} we plot  the structure of  the adjoint in the energy inner product of the unstable KH  mode   shown in  FIG.~\ref{fig:KH_mode_Ri0_0p06} with $k=1$, $\Ri(0)=0.06$ and $R=3$.
An initial condition consisting of the adjoint of the most unstable mode   
excites the mode with energy 7.5 times greater than an initial condition 
consisting of the unstable  mode itself. 
In FIGs.~\ref{fig:Holmboe_mode1_adjoints} and \ref{fig:Holmboe_mode2_adjoints}  we plot  the adjoints  of the  two  unstable Holmboe H1 and H2 modes, shown in FIGs.~\ref{fig:Holmboe_mode1} and \ref{fig:Holmboe_mode2}, both with the same $k=1$, $R=3$ and stratifications $\Ri(0)=0.65$ and $\Ri(0)=12$ respectively. 
The adjoints of the H1 and H2  modes are concentrated at the critical layer of the modes which is located way from the center of the shear layer and located in a region at the wings of the shear layer where the local Richardson number is less than $1/4$. 
An initial condition
consisting of the adjoint of the H1 mode excites the most unstable mode with energy  69 greater  than an initial condition consisting of the unstable  mode itself. This  energy amplification reaches  7921 for the excitation of the H2 mode by its adjoint.

Initial conditions in the form of the adjoints  of the most unstable mode evolve changing form and eventually  assume the corresponding modal form.
During their evolution,  the adjoint perturbation extracts energy from the mean flow  which is eventually deposited to the mode itself exciting it in this way at high amplitude.
The time development of the energy of a unit energy initial condition in the form of the adjoints of the H1 and H2 mode is plotted in FIG.~\ref{fig:H1H2adjt} demonstrating the increased excitation of the corresponding modes. 
This demonstrates that the  unstable Holmboe modes at high Richardson numbers arise primarily due to excitation by their adjoint and the modal growth underestimates the growth 
of the instabilities.

\section{\label{sec:nonnormal_finite}Optimal growth of perturbations and the emergence of the Holmboe quasi-mode}

In the previous section we have demonstrated that for large Richardson numbers the adjoint of a mode can excite the modes at much higher amplitude.
In this section we investigate the optimal growth of perturbations. The optimal growth\cite{Farrell-1988a, Farrell-Ioannou-1996a,Schmid-Henningson-2001} of perturbations at time $t$ in the energy metric is obtained by calculating the norm of the  propagator $e^{\A_{\M} t}$  where $\A_{\M}= \M^{1/2}\A\M^{-1/2}$ and $\M$ is the energy metric defined as
\be
\M = \frac{\D z}{4} 
\[
\begin{array}{cc}
-(\Df^2-k^2)  & 0    \\
  0    & -J \rho_0'   
\end{array}
\]~,
\ee
so that total perturbation energy is given by $E=\hat \x^\dag \M \hat{\x}$. The  optimal growth is given by the largest singular value, $\sigma_{\max}$, of  $e^{\A_{\M} t}$ and determines the largest perturbation growth that can be achieved at time $t$. The optimal perturbation, the initial perturbation that produces this growth, is $\hat \x_{\text{opt}} = \M^{-1/2} \v$, where $\v$ the right singular vector with singular value $\sigma_{\max}$.

\begin{figure*}
\includegraphics{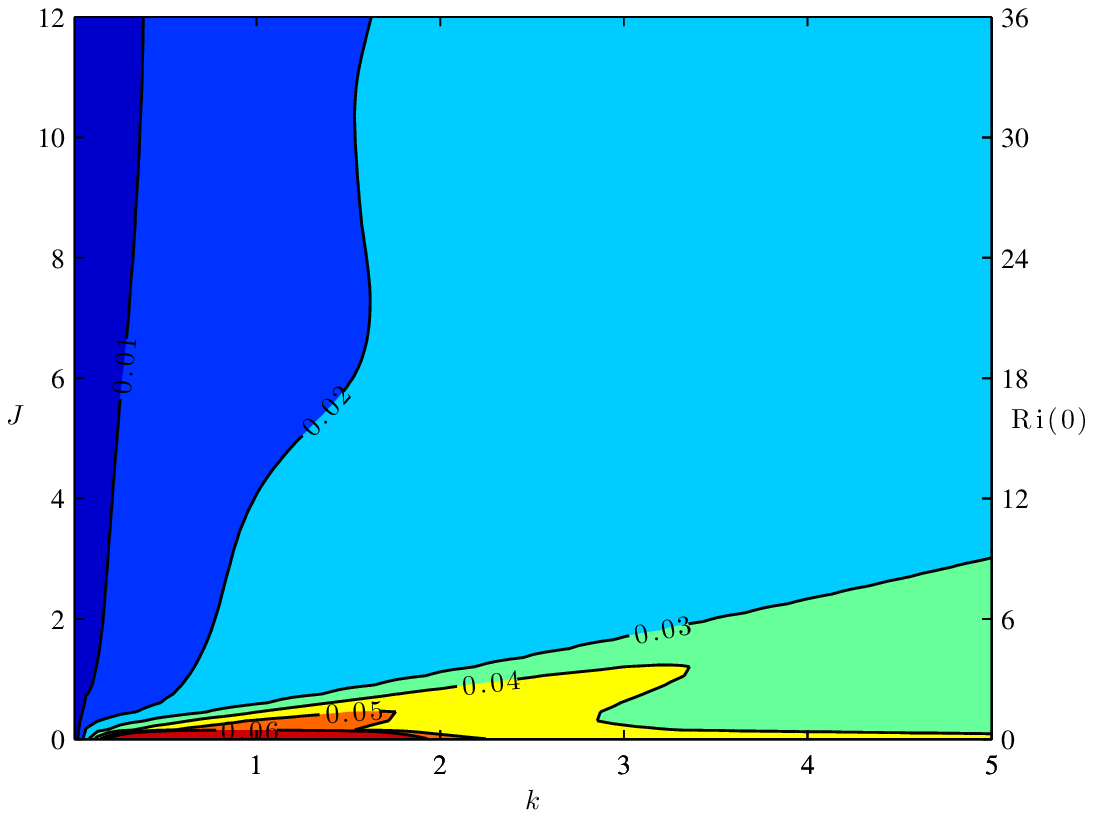}
\caption{\label{fig:growth_R_3_Topt_100}(Color online) Optimal growth for $t=100$. 
Contours of  finite time Lyapunov exponent $\ln{\(\s_{\max}(t)\)}/t$ for  $t=100$ as a function of wavenumber, $k$, and the bulk Richardson number, $J$, for $R=3$. 
The corresponding value of the Richardson number at the origin $\Ri(0)=R J $ is indicated  at the ordinate axis on the right.}
\end{figure*}

\begin{figure*}
\includegraphics{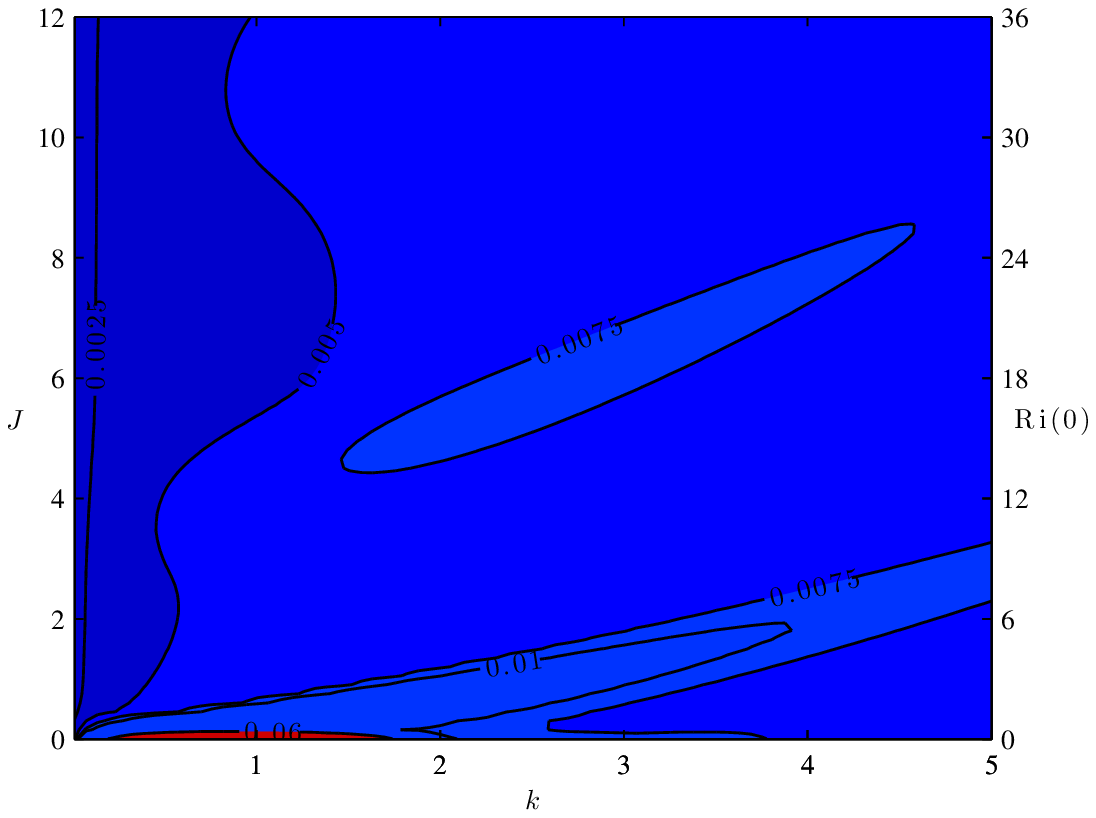}
\caption{\label{fig:growth_R_3_Topt_600}(Color online)  Optimal growth for $t=600$. Contours of finite time Lyapunov exponent $\ln{\(\s_{\max}(t)\)}/t$ for  $t=600$ as a function of wavenumber, $k$, and the bulk Richardson number, $J$, for $R=3$. The corresponding value of the Richardson number at the origin $\Ri(0)=R J $ is indicated  at the ordinate axis on the right.}
\end{figure*}

\begin{figure*}
\includegraphics{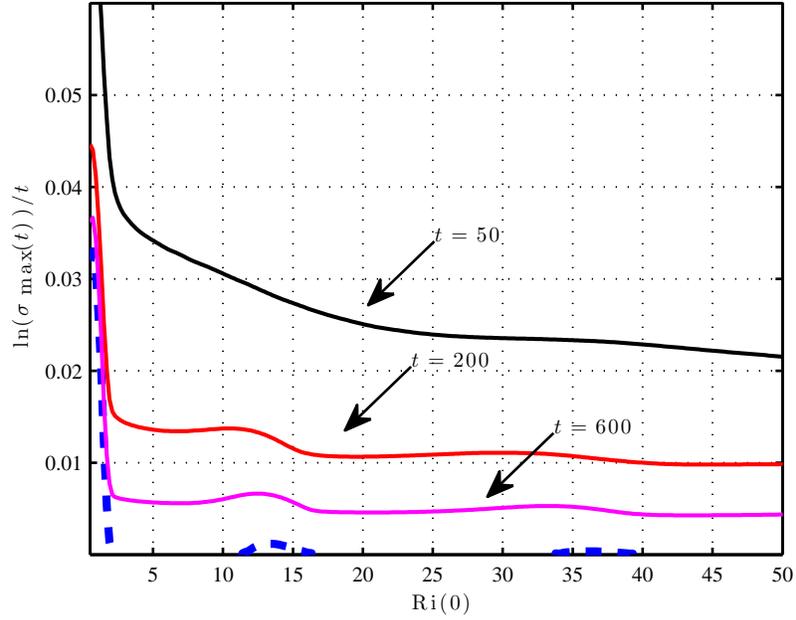}
\caption{\label{fig:optkcivsRi}(Color online) 
Comparison of the finite time Lyapunov exponent, $\ln{\(\s_{\max}(t)\)}/t$ (solid), associated with the growth of optimal perturbations for optimizing times $t=50$, $t=200$ and $t=600$, 
with  the modal growth rate $kc_i$  (dashed) for various Richardson numbers.
The first three unstable Holmboe branches are shown. The  H1 branch is for  $\Ri(0)<1.42$.
The H2 branch  is for  $10.46\le\Ri(0)\le 16.26$; and  H3 is for $31.04\le\Ri(0)\le 39.63$. 
The optimal growth rate is almost constant for large Richardson numbers  and is continuous across the islands of instability. 
Parameters are:  $k=1$ and  $R=3$.}
\end{figure*}

We calculate the optimal growth for two indicative optimizing times $t=100$ and $t=600$ and plot the finite time Lyapunov exponent $\ln{\(\s_{\max}(t)\)}/t$ associated with the optimal perturbations as a function of wavenumber $k$ and stratification $J$. As $t \rightarrow \infty$  the exponent tends to the modal growth rate $\ln{\(\s_{\max}(t)\)}/t \rightarrow kc_i$. We saw that the modal growth rate (cf. FIG.~\ref{fig:spectrum_R_3_LARGE}) is non zero only in narrow bands of parameter space.
In contrast, the growth rate associated with the optimal perturbations covers all parameter space. For example, for optimizing time $t=100$ the finite Lyapunov exponent, shown in FIG.~\ref{fig:growth_R_3_Topt_100},  is almost constant for large Richardson numbers and the equivalent growth rates for large Richardson numbers are at least an order of magnitude larger than the modal growth rates for all values of the parameters. Furthermore the growth rates do not reveal the underlying bands of exponential instability. 
The optimal growth is robust even for optimizing time $t=600$. Contours of the finite Lyapunov exponents for this optimizing time (FIG.~\ref{fig:growth_R_3_Topt_600}) show  that the optimal  growth rates continue to be at least an order of magnitude larger than the modal growth rate for all parameter values.
Moreover the optimal growth rates are  almost constant as a function of Richardson number  for large  Richardson numbers. 
This can be seen in FIG.~\ref{fig:optkcivsRi} where we compare the optimal growth rate for optimizing times $t=50,\ 200,\ 600$ with the  modal growth rate as a function of the stratification at the center $\Ri(0)$ for $k=1$.  While the modal growth rate is only substantial for the H1 branch of instability the optimal growth rates produce sustainable growth at all stratifications and  is reduced  by a factor of  less than 2 when the center Richardson number increases tenfold from 5 to 50. 

\begin{figure*}
\includegraphics{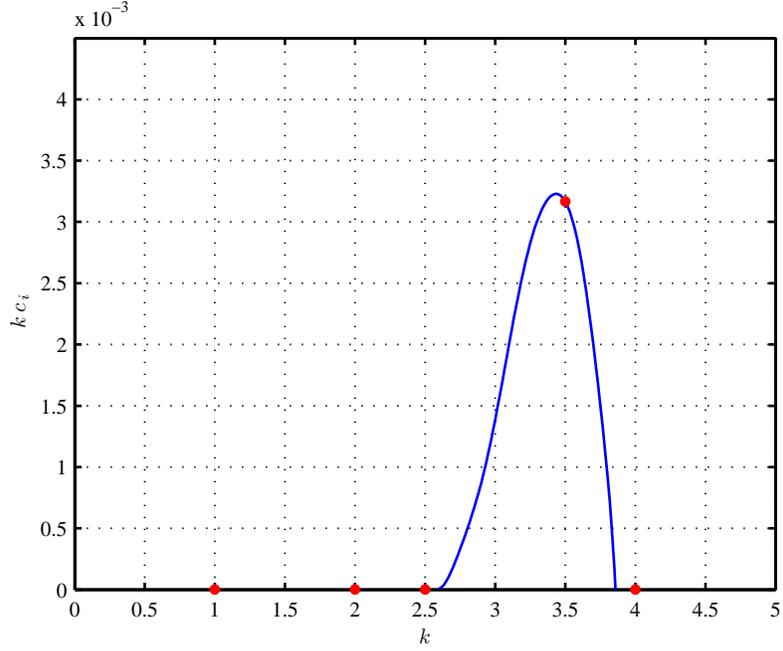}
\caption{\label{fig:growth_rates_wavenumber_N_501_R_3_Ri0_5}(Color online) Modal growth rate $kc_i$ as a function of wavenumber $k$ for center Richardson number $\Ri(0)=5$. 
The instability for $2.6<k<3.9$ corresponds to the  H1 branch. 
The dots indicate the growth rate at $k=1,\ 2,\ 2.5,\ 4$ and at $k=3.5$ when  the maximum growth rate occurs.}
\end{figure*}

\begin{figure*}
\includegraphics{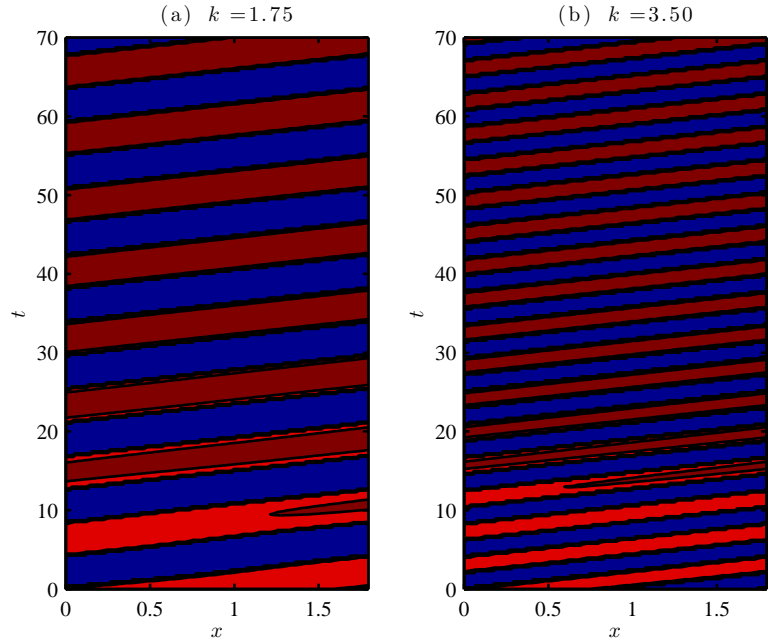}
\caption{\label{fig:hovmoller}(Color online) Contours of the logarithm of the positive real part of the perturbation $\Re(\hat \psi(z,t) e^{ i k x})$ at $z=0.7$ in the $(x,t)$ plane for an initial perturbation in the form of the adjoint in the energy
norm of the most unstable Holmboe mode at $k=3.5$ for $R=3$ and $\Ri(0)=5$.
Panel (a): evolution under the dynamics with $k=1.75$ when the flow is stable.  Panel (b): evolution under the dynamics with  $k=3.5$ when the flow is unstable and this adjoint excites optimally the prograde Holmboe mode. The same propagation characteristics emerge at other wavenumbers for which the flow is stable. 
These  Hovm\"oller diagrams indicate that when the flow is stable a quasi-mode emerges propagating with  phase speed close to that of  the Holmboe mode.}
\end{figure*}

\begin{figure*}
\includegraphics{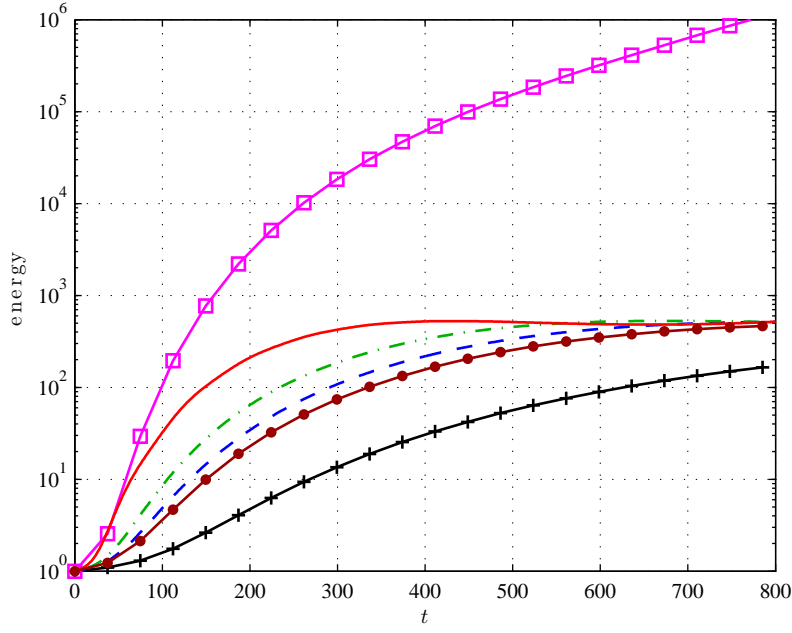}
\caption{\label{fig:Ekadj}(Color online) Energy evolution of a unit
energy initial condition with the structure of the adjoint in the energy norm
of the unstable H1 mode
at  $k=3.5$, $R=3$ and $\Ri(0)=5$ with the dynamical operator
(a) for the unstable $k=3.5$ (solid-squares), (b) 
for $k=1.0$ (solid crosses), $k=1.75$
(solid-dots), $k=2.0$ (dashed),  $k=2.5$ (dash-dot) and  $k=4.0$
(solid) for which the flow is stable. 
When the flow is stable large
amplitude propagating quasi-modes emerge as shown in
FIG.~\ref{fig:hovmoller}.}
\end{figure*}

The structure of the optimal perturbations for the larger optimizing times is very close to the structure of the adjoint of the modes, even for wavenumbers for which there is no Holmboe instability. 
This reveals that excitation of these flows at high Richardson numbers will lead to the emergence of propagating perturbations which are close in structure to  Holmboe waves.  To be specific, consider central stratification $\Ri(0)=5$ and excitation of the shear layer with various wavenumbers $k$. In FIG.~\ref{fig:growth_rates_wavenumber_N_501_R_3_Ri0_5} we plot the growth rate as a function of $k$. There is a band of instability  producing H1 waves with phase speeds close to $c_r  \approx \pm 0.5$ and when the flow is excited at these wavenumbers  pairs of propagating waves emerge. This can be seen in 
the Hovm\"oller diagram FIG.~\ref{fig:hovmoller}(b) in which contours of the logarithm of the positive real part of $\Re( \hat \psi(z,t) e^{ i k x})$ for $k=3.5$ and fixed $z=0.7$ are plotted in the $(x,t)$ plane for an  initial condition in the form of  the adjoint of the unstable H1 mode.
The characteristics of the  prograde  propagating Holmboe wave emerge from the start.  In FIG.~\ref{fig:hovmoller}(a) we plot the corresponding Hovm\"oller diagram when the same  adjoint is introduced as initial condition and evolved this time with the dynamics for $k=1.75$. Although at  $k=1.75$ there is no instability, a propagating structure emerges with the characteristics of the unstable Holmboe wave. The same propagating structure also emerges at other wavenumbers $k$. Therefore the Holmboe wave is a robust dynamic entity that forms a quasi-mode at wavenumbers which do not support an unstable Holmboe wave. This quasi-mode is the manifestation of the near resonant edge wave structures that lead to the Holmboe instability at resonance\cite{Holmboe-1962,Baines-Mitsudera-1994,Caulfield-1994,Haigh-Lawrence-1999,Harnik-etal-2008,Carpenter-etal-2010}. 
Similar  quasi-mode behavior is seen in the evolution of the edge waves
in the Holmboe profile (cf. Appendix~\ref{sec:Holmboe_classical}) for  wavenumbers $k$ for which there is no instability.
When  $k$ is close to the  wavenumber for which there is instability 
the edge wave complex  periodically amplifies
while propagating at a phase speed close to the
phase speed  of the unstable Holmboe modes. 
What is surprising here is that  these propagating quasi-modes  can be excited 
at such high amplitude and that their
growth is so persistent. 
This is shown in FIG.~\ref{fig:Ekadj} where the adjoint of the Homboe instability for $k=3.5$  excites at high amplitude  quasi-modes for the wavenumbers  $k= 1,\ 2,\ 2.5,\ 4$ for which the flow is stable. The same highly amplified quasi-mode would emerge  if we initialize the flow with a large time optimal.

\begin{figure*}
\includegraphics{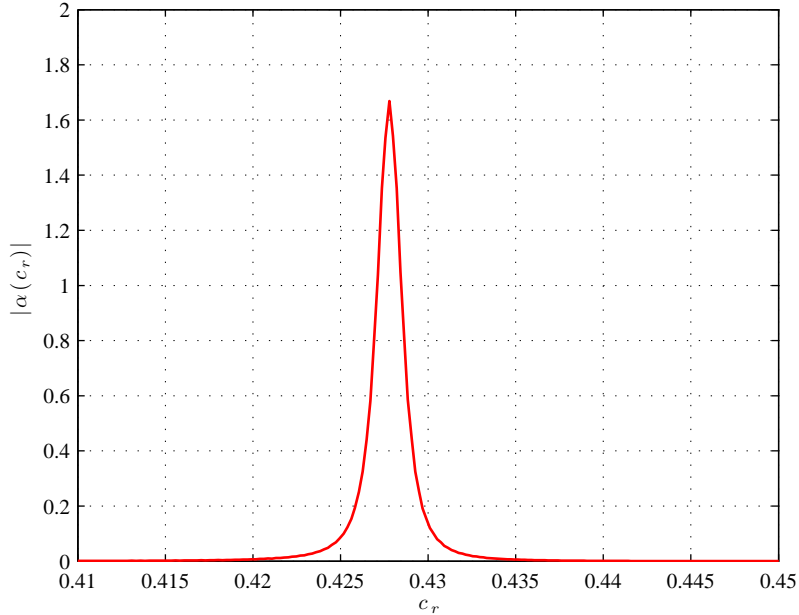}
\caption{\label{fig:Fk2}(Color online) Amplitude, $|\alpha(c_r)|$, of
the absolute value of the  expansion coefficients of the quasi-mode
at $k=1.75$ as a function of the phase speed of the modes of the perturbation operator at $k=1.75$. 
All the  modes of the perturbation operator are
stable.
The quasi-mode is excited by a unit energy initial condition with
the structure of the adjoint of the
unstable mode for $k=3.5$.
Notice the sharp peak at $c_r \approx0.4278$ which indicates
the formation of long-lived quasi-modes that propagate with this phase
speed (see FIG.~\ref{fig:hovmoller}(a)). For $R=3$ and $\Ri(0)=5$.}
\end{figure*}

The structure of the quasi-mode can not be ascribed to the structure of any single
mode of the operator. The quasi-mode is the superposition of a multitude of
continuum spectrum modes.  The quasi-mode can propagate as a coherent entity at a single
phase speed because the distribution of the amplitude of the coefficients of the modal expansion is sharply
peaked at the phase speed of the propagation of the quasi-mode (the amplitude of the coefficients
is time invariant because all modes have zero growth). For example,
the distribution of
the amplitude of coefficients of the modal expansion of the quasi-mode  for $k=1.75$
in FIG.~\ref{fig:Fk2}, is concentrated at the phase speed $c_r \approx \pm 0.4278$  which is
exactly the phase speed that emerges in
FIG.~\ref{fig:hovmoller}(a).

\begin{figure*}
\includegraphics{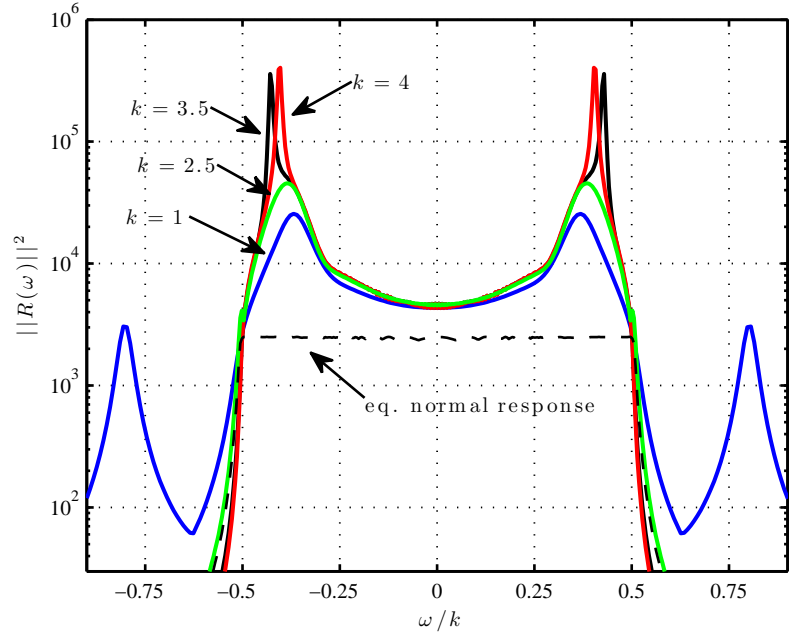}
\caption{\label{fig:Romega}(Color online) The maximum energy response,  $||\R(\omega)||^2$,  to harmonic forcing as a function of $\omega / k$ for perturbations with $k=1$, $k=2.5$, $k=3.5$ and $k=4.0$. The dynamics have been rendered stable by addition of Rayleigh friction. The response is maximized at the phase speed of the
quas-mode. 
Also shown is the  resonant frequency response that would obtain if the operators were normal (dashed). The 
amplified response in energy is a measure of the non-normality of the operator. For stratification $\Ri(0)=5$ and $R=3$.}
\end{figure*}

The quasi-mode can be identified from the   frequency response of the perturbation dynamics 
to harmonic forcing. Because the operators are either neutral or unstable in order to obtain steady harmonic response we introduce a linear damping  in the dynamics, that does not affect the eigenstructures, and consider the frequency response of the stable operators $\tilde \A_{\M}=\A_{\M} -0.02\,\1$, where $\1$ is the identity.
The steady state harmonic response, $\tilde \psi(z,\omega) $, where $\hat \psi (z,t) = \tilde{\psi}(z,\omega) e^{i \omega t}$, to harmonic forcing  $\F e^{i \omega t}$ is then given by
\begin{equation}
\tilde \psi(z,\omega)  = \R_k( \omega) \F~,
\end{equation}
where $\R_k(\omega)$ is the resolvent given by
\begin{equation}
\R_k (\omega) = (i \omega \1 - \tilde \A_{\M})^{-1}~.
\end{equation}
The maximum possible response to harmonic forcing is then given by the square of the 2-norm of the resolvent, $||\R_k (\omega)||^2$, which is equal to the square of its largest singular value. This determines  the maximum energy that can be achieved at frequency $\omega$ by unit energy harmonic excitation. The frequency response shown in FIG.~\ref{fig:Romega} demonstrates the concentration of the response at the phase speeds of the emergent quasi-modes which is very close to the phase speed of the Holmboe waves at $k=3.5$. 
It demonstrates also the importance of the non-normal interactions in the excitation at high amplitude of the quasi-modes. In the same plot we plot the  equivalent normal frequency response
\begin{equation}
\max_j \frac{1}{| i \omega - i k c_j |^2}~,
\label{eq:normal_freq_response}
\end{equation}
where $i k c_j$ is the $j$-th eigenvalue of $\A_{\M}$ that would have resulted if the eigenvectors were orthogonal.  The difference between the responses reflects the excess energy that is maintained by the system against friction because of the non-orthogonality of the eigenmodes\cite{Farrell-Ioannou-1994b, Ioannou-1995}.

\section{Conclusions}

Highly stratified shear layers are  susceptible to Holmboe instabilities which can lead to mixing.
We have demonstrated in this work that especially at large Richardson  numbers the adjoint of the  weakly unstable modes can excite, through potent transfer of energy from the mean flow, the unstable  modes at high amplitude. Further we found, that even in regions of parameter space where the flow is neutral, optimal perturbations can grow strongly and excite at large amplitude long-lived propagating quasi-modes. 
We have demonstrated  that the modal growth substantially underestimates the growth potential  of perturbations in
such highly stratified shear layers.

\acknowledgments
Navid Constantinou gratefully acknowledges  the partial support of  the A. G. Leventis Foundation and J. S. Latsis Public Benefit Foundation.

\appendix
\section{\label{sec:Holmboe_classical}Non-modal growth produced by  the edge waves in the idealized Holmboe background state}

\begin{figure*}
\includegraphics{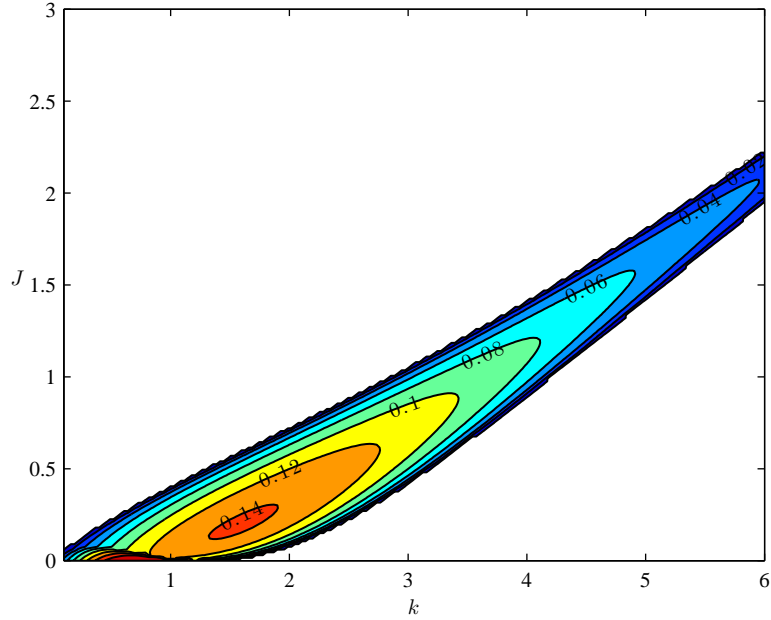}
\caption{\label{fig:spectrum_edgewave}(Color online) Contours of modal growth rate $kc_i$ 
as a function of wavenumber, $k$, and stratification parameter, $J$, of perturbations in Holmboe's idealized piecewise constant background state. 
At larger values of $k$ and $J$ there is a band of instability associated with the Holmboe instability.}
\end{figure*}

We consider perturbations in Holmboe's\cite{Holmboe-1962} idealized piecewise linear velocity profile $U_0(z)=z$ for $|z|\le 1/2$ and $U_0(z)=z/(2|z|)$ for $|z|>1/2$  and mean density $\rho_0(z)=-z/|z|$ in an infinite domain.
This profile has two vorticity discontinuities at $z=\pm 1/2$ and a density discontinuity at $z=0$; each vorticity discontinuity supports
an edge wave and the density discontinuity supports a pair of edge waves one prograde and the other retrograde~\cite{Holmboe-1962, Baines-Mitsudera-1994,Harnik-etal-2008,Rabinovich-etal-2011}. 
Consider perturbations with streamfunction $\hat{\psi}(z,t) e^{ikx}$. A time dependent solution to the perturbation equations can be obtained\cite{Holmboe-1962, Baines-Mitsudera-1994} by introducing   solutions of the form, 
\begin{subequations}
\label{eq:psi_solutions}
\begin{align}
\hat{\psi}_1(z,t) &= A_1(t) e^{-k(z-1/2)}, \quad{\rm for} ~~z>1/2\\
\hat{\psi}_2(z,t) &= A_2(t) e^{-kz} +  B_2(t) e^{kz}, \quad{\rm for}~~ 0<z <1/2\\
\hat{\psi}_3(z,t) &= A_3(t) e^{-kz} +  B_3(t) e^{kz}, \quad{\rm for}~~ -1/2<z <0\\
\hat{\psi}_4(z,t) &= B_4(t) e^{k(z+1/2)}~, \quad{\rm for}~~ 0<z <1/2
\end{align}
\end{subequations}
which  reduce them, by imposing
the usual continuity conditions at the discontinuity interfaces,
to a set of ordinary differential equations for the evolution of the amplitudes $A$ and $B$. We choose as a state variable
$\u(t)=\[A_1(t),C_2(t),B_4(t),\eta_0(t)\]^\transp$,  where $\eta_0(t)\equiv \hat{\eta}(z=0,t)$ is the displacement at $z=0$
and $C_2(t) \equiv \(A_2(t) + B_2(t)\)e^{k/2}$.  The perturbation equations are thus reduced to:
\be
\frac{\df \u }{\df t}  = \A \, \u~~,
\label{eq:system_analytical}
\ee
with $\A = \K^{-1} \L$ with the matrices $\K$ and $\L$ given by:
\begin{widetext}
\begin{subequations}
\begin{align}
\K&= \[
\begin{array}{cccc}
  -e^{k}	& 1				& 0		& 0   \\
  2		&-2\(1+e^{-k}\)		& 2		&0   \\
  0		&1				&-e^{k}	& 0\\
  0		& 0				&0		&1   
\end{array}
\]~~,
 \\
\L &= i\[
\begin{array}{cccc}
   \(1+e^{k}(k-1)\)/2		& -k/2 		& 0					& 0   \\
  0					&0			& 0					& J  \(e^{k/2}-e^{-k/2}\)	 \\
  0					&k/2			&-\(1+e^{k}(k-1)\)/2		& 0\\
  0					& -k e^{-k/2}	&0					&0   
\end{array}
\]~.
\end{align}
\end{subequations}
\end{widetext}
These equations determine the time evolution of the perturbation structure as determined by the interaction of the four edge waves. It determines fully
the modal stability properties of the background state and the non-normal dynamics that derive from the interaction of the edge waves.  
In this formulation we have neglected the  continuum spectrum. A contour plot of the resulting growth rate is shown in
FIG.~\ref{fig:spectrum_edgewave} 
as a function of wavenumber, $k$, and stratification parameters, $J$. The KH branch of instability is at low $k$ and $J$
and the narrowing band of the Holmboe mode of instability at larger $k$ and $J$. 

In order to study the non-normal dynamics associated with the edge waves we must introduce the  corresponding to  \eqref{eq:def_norm} energy metric. 
The potential energy, since $\rho_0' = -\delta(z)$,  is $P = \frac1{4} J\, \left|\eta_0\right|^2$. The kinetic energy can be written as:
\begin{widetext}
\be
\begin{split}
T = &\frac1{4}\times\( \left. -\hat{\psi_1}^*\,\Df \hat{\psi_1} \right|_{z=1/2} +\left.\hat{\psi_2}^*\,\Df \hat{\psi_2}   \right|_{z=1/2} -\left.\hat{\psi_2}^*\,\Df \hat{\psi_2}   \right|_{z=0} \)\\
&+\frac1{4}\times\( \left.\hat{\psi_3}^*\,\Df \hat{\psi_3}   \right|_{z=0} -\left. \hat{\psi_3}^*\,\Df \hat{\psi_3}   \right|_{z=-1/2} +\left.\hat{\psi_4}^*\,\Df \hat{\psi_4}   \right|_{z=-1/2}\) ~.
\end{split}
\ee
\end{widetext}
We  thus derive that the perturbation  energy of the state $\u(t)$  is given by
\be
E=\u^{\dagger} \M \u
\ee
with the metric, $\M$, given by 
\be
\M= \frac1{4}\[
\begin{array}{cc}
  \M_{\textrm{T}}	& 0   \\
  0			& J	   
\end{array}
\]
\ee
where $\M_{\textrm{T}}$ is:
\be
\M_{\textrm{T}}= \frac{2k}{\(e^k-1\)}\[
\begin{array}{ccc}
  e^{k}		& -1			& 0   \\
  -1		&1+e^{-k}		&-1   \\
  0		&-1			& e^k   
\end{array}
\]
\ee

\begin{figure*}
\includegraphics{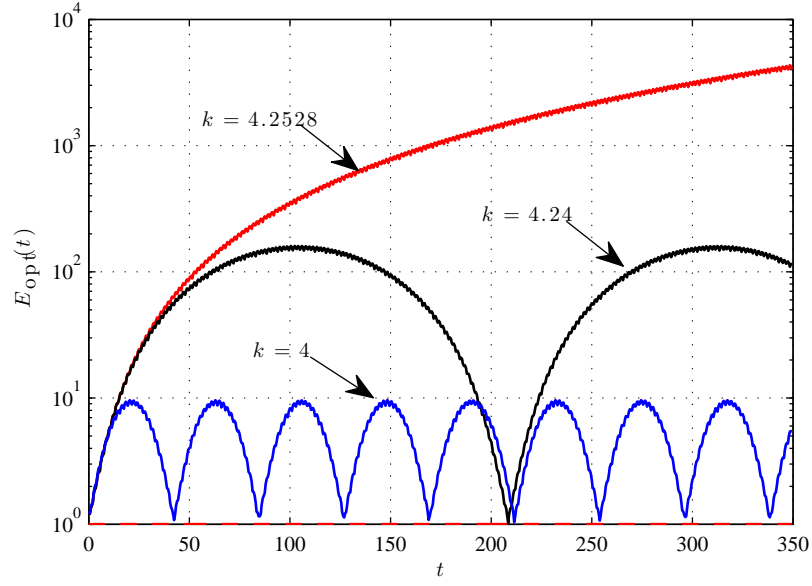}
\caption{\label{fig:Eholmboe}(Color online) Optimal energy growth that can be produced by the edge waves 
as a function of time for stratification parameter $J=1.5$ for 
$k=4.2528,~4.24,~4$. All $k\le 4.2528$  are modally stable ($kc_i=0$) and  $k=4.2528$ is located at the stability boundary.
The non-modal growth produced by the edge waves is substantial as the stability boundary is approached. } 
\end{figure*}

The non-modal growth associated by  the dynamics of the edge waves is obtained by calculating the
optimal energy growth that   can be achieved at time $t$ which is given by:
\be
E_{\textrm{opt}}(t) = || \exp (\A_{\textrm{M}} t)||^2
\ee
where $\A_{\textrm{M}} =  \M^{1/2} \A \M^{-1/2}$.  It can be shown that the edge waves can lead to substantial growth for parameter
values for which there is no instability but are close to the stability boundary. For example
consider the large value of stratification $J=1.5$. 
The flow supports unstable waves  only for $4.2528<k<  5.142$ and the  
maximum growth rate is $0.064$.
The optimal energy growth produced by the edge waves for $k=4.2528,~4.24,~4$  is shown in FIG.~\ref{fig:Eholmboe}
to be substantial as the stability boundary is approached.


\begin{thebibliography}{58}

\expandafter\ifx\csname natexlab\endcsname\relax\def\natexlab#1{#1}\fi
\expandafter\ifx\csname bibnamefont\endcsname\relax
  \def\bibnamefont#1{#1}\fi
\expandafter\ifx\csname bibfnamefont\endcsname\relax
  \def\bibfnamefont#1{#1}\fi
\expandafter\ifx\csname citenamefont\endcsname\relax
  \def\citenamefont#1{#1}\fi
\expandafter\ifx\csname url\endcsname\relax
  \def\url#1{\texttt{#1}}\fi
\expandafter\ifx\csname urlprefix\endcsname\relax\def\urlprefix{URL }\fi
\providecommand{\bibinfo}[2]{#2}
\providecommand{\eprint}[2][]{\url{#2}}


\bibitem[{\citenamefont{Miles}(1961)}]{Miles-1961}
\bibinfo{author}{\bibfnamefont{J.~W.} \bibnamefont{Miles}},
  \bibinfo{journal}{J. Fluid Mech.} \textbf{\bibinfo{volume}{10}},
  \bibinfo{pages}{496} (\bibinfo{year}{1961}).

\bibitem[{\citenamefont{Howard}(1961)}]{Howard-1961}
\bibinfo{author}{\bibfnamefont{L.~N.} \bibnamefont{Howard}},
  \bibinfo{journal}{J. Fluid Mech.} \textbf{\bibinfo{volume}{10}},
  \bibinfo{pages}{509} (\bibinfo{year}{1961}).

\bibitem[{\citenamefont{Drazin and Reid}(1981)}]{Drazin-Reid-81}
\bibinfo{author}{\bibfnamefont{P.~G.} \bibnamefont{Drazin}} \bibnamefont{and}
  \bibinfo{author}{\bibfnamefont{W.~H.} \bibnamefont{Reid}},
  \emph{\bibinfo{title}{Hydrodynamic Stability}} (\bibinfo{publisher}{Cambridge
  University Press, Cambridge}, \bibinfo{year}{1981}).

\bibitem[{\citenamefont{Holmboe}(1962)}]{Holmboe-1962}
\bibinfo{author}{\bibfnamefont{J.}~\bibnamefont{Holmboe}},
  \bibinfo{journal}{Geophys. Publ.} \textbf{\bibinfo{volume}{24}},
  \bibinfo{pages}{67} (\bibinfo{year}{1962}).

\bibitem[{\citenamefont{Thorpe}(1971)}]{Thorpe-1971}
\bibinfo{author}{\bibfnamefont{S.~A.} \bibnamefont{Thorpe}},
  \bibinfo{journal}{J. Fluid Mech.} \textbf{\bibinfo{volume}{46}},
  \bibinfo{pages}{299} (\bibinfo{year}{1971}).

\bibitem[{\citenamefont{Browand and Winant}(1973)}]{Browand-Winant-1973}
\bibinfo{author}{\bibfnamefont{F.~K.} \bibnamefont{Browand}} \bibnamefont{and}
  \bibinfo{author}{\bibfnamefont{C.}~\bibnamefont{Winant}},
  \bibinfo{journal}{Boundary Layer Meteorol.} \textbf{\bibinfo{volume}{5}},
  \bibinfo{pages}{67} (\bibinfo{year}{1973}).

\bibitem[{\citenamefont{Pouliquen et~al.}(1994)\citenamefont{Pouliquen, Chomaz,
  and Huere}}]{Pouliquen-etal-1994}
\bibinfo{author}{\bibfnamefont{O.}~\bibnamefont{Pouliquen}},
  \bibinfo{author}{\bibfnamefont{J.-M.} \bibnamefont{Chomaz}},
  \bibnamefont{and} \bibinfo{author}{\bibfnamefont{P.}~\bibnamefont{Huere}},
  \bibinfo{journal}{J. Fluid Mech.} \textbf{\bibinfo{volume}{266}},
  \bibinfo{pages}{277} (\bibinfo{year}{1994}).

\bibitem[{\citenamefont{Caulfield et~al.}(1995)\citenamefont{Caulfield,
  Peltier, Yoshida, and Ohtani}}]{Caulfield-etal-1995}
\bibinfo{author}{\bibfnamefont{C.~P.} \bibnamefont{Caulfield}},
  \bibinfo{author}{\bibfnamefont{W.~R.} \bibnamefont{Peltier}},
  \bibinfo{author}{\bibfnamefont{S.}~\bibnamefont{Yoshida}}, \bibnamefont{and}
  \bibinfo{author}{\bibfnamefont{M.}~\bibnamefont{Ohtani}},
  \bibinfo{journal}{Phys. Fluids} \textbf{\bibinfo{volume}{7}},
  \bibinfo{pages}{3028} (\bibinfo{year}{1995}).

\bibitem[{\citenamefont{Zhu and Lawrence}(2001)}]{Zhu-Lawrence-2001}
\bibinfo{author}{\bibfnamefont{D.~Z.} \bibnamefont{Zhu}} \bibnamefont{and}
  \bibinfo{author}{\bibfnamefont{G.~A.} \bibnamefont{Lawrence}},
  \bibinfo{journal}{J. Fluid Mech.} \textbf{\bibinfo{volume}{429}}
  (\bibinfo{year}{2001}).

\bibitem[{\citenamefont{Hogg and Ivey}(2003)}]{Hogg-Ivey-2003}
\bibinfo{author}{\bibfnamefont{A.}~\bibnamefont{Hogg}} \bibnamefont{and}
  \bibinfo{author}{\bibfnamefont{G.}~\bibnamefont{Ivey}}, \bibinfo{journal}{J.
  Fluid Mech.} \textbf{\bibinfo{volume}{477}}, \bibinfo{pages}{339}
  (\bibinfo{year}{2003}).

\bibitem[{\citenamefont{Tedford et~al.}(2009)\citenamefont{Tedford, Pieters,
  and Lawrence}}]{Tedford-etal-2009}
\bibinfo{author}{\bibfnamefont{E.~W.} \bibnamefont{Tedford}},
  \bibinfo{author}{\bibfnamefont{R.}~\bibnamefont{Pieters}}, \bibnamefont{and}
  \bibinfo{author}{\bibfnamefont{G.~A.} \bibnamefont{Lawrence}},
  \bibinfo{journal}{J. Fluid Mech.} \textbf{\bibinfo{volume}{636}},
  \bibinfo{pages}{137} (\bibinfo{year}{2009}).

\bibitem[{\citenamefont{Smyth et~al.}(1988)\citenamefont{Smyth, Klaassen, and
  Peltier}}]{Smyth-etal-1988}
\bibinfo{author}{\bibfnamefont{W.~D.} \bibnamefont{Smyth}},
  \bibinfo{author}{\bibfnamefont{G.~P.} \bibnamefont{Klaassen}},
  \bibnamefont{and} \bibinfo{author}{\bibfnamefont{W.~R.}
  \bibnamefont{Peltier}}, \bibinfo{journal}{Geophys. Astrophys. Fluid Dyn.}
  \textbf{\bibinfo{volume}{43}}, \bibinfo{pages}{181} (\bibinfo{year}{1988}).

\bibitem[{\citenamefont{Sutherland et~al.}(1994)\citenamefont{Sutherland,
  Caulfield, and Peltier}}]{Sutherland-etal-1994}
\bibinfo{author}{\bibfnamefont{B.~R.} \bibnamefont{Sutherland}},
  \bibinfo{author}{\bibfnamefont{C.~P.} \bibnamefont{Caulfield}},
  \bibnamefont{and} \bibinfo{author}{\bibfnamefont{W.~R.}
  \bibnamefont{Peltier}}, \bibinfo{journal}{J. Atmos. Sci.}
  \textbf{\bibinfo{volume}{51}}, \bibinfo{pages}{3261} (\bibinfo{year}{1994}).

\bibitem[{\citenamefont{Smyth}(2006)}]{Smyth-2006}
\bibinfo{author}{\bibfnamefont{W.~D.} \bibnamefont{Smyth}},
  \bibinfo{journal}{Phys. Fluids} \textbf{\bibinfo{volume}{18}},
  \bibinfo{pages}{064104} (\bibinfo{year}{2006}).

\bibitem[{\citenamefont{Smyth and Peltier}(2006)}]{Smyth-Peltier-2006}
\bibinfo{author}{\bibfnamefont{W.~D.} \bibnamefont{Smyth}} \bibnamefont{and}
  \bibinfo{author}{\bibfnamefont{W.~R.} \bibnamefont{Peltier}},
  \bibinfo{journal}{J. Fluid Mech.} \textbf{\bibinfo{volume}{228}},
  \bibinfo{pages}{387} (\bibinfo{year}{2006}).

\bibitem[{\citenamefont{Smyth et~al.}(2007)\citenamefont{Smyth, Carpenter, and
  Lawrence}}]{Smyth-etal-2007}
\bibinfo{author}{\bibfnamefont{W.~D.} \bibnamefont{Smyth}},
  \bibinfo{author}{\bibfnamefont{J.~R.} \bibnamefont{Carpenter}},
  \bibnamefont{and} \bibinfo{author}{\bibfnamefont{G.~A.}
  \bibnamefont{Lawrence}}, \bibinfo{journal}{J. Atmos. Sci.}
  \textbf{\bibinfo{volume}{37}}, \bibinfo{pages}{1566} (\bibinfo{year}{2007}).

\bibitem[{\citenamefont{Carpenter et~al.}(2007)\citenamefont{Carpenter,
  Lawrence, and Smyth}}]{Carpenter-etal-2007}
\bibinfo{author}{\bibfnamefont{J.~R.} \bibnamefont{Carpenter}},
  \bibinfo{author}{\bibfnamefont{G.~A.} \bibnamefont{Lawrence}},
  \bibnamefont{and} \bibinfo{author}{\bibfnamefont{W.~D.} \bibnamefont{Smyth}},
  \bibinfo{journal}{J. Fluid Mech.} \textbf{\bibinfo{volume}{582}},
  \bibinfo{pages}{103} (\bibinfo{year}{2007}).

\bibitem[{\citenamefont{Alexakis}(2009)}]{Alexakis-2009}
\bibinfo{author}{\bibfnamefont{A.}~\bibnamefont{Alexakis}},
  \bibinfo{journal}{Phys. Fluids} \textbf{\bibinfo{volume}{21}},
  \bibinfo{pages}{054108} (\bibinfo{year}{2009}).

\bibitem[{\citenamefont{Tedford}(2009)}]{Tedford-2009}
\bibinfo{author}{\bibfnamefont{E.~D.} \bibnamefont{Tedford}}, Ph.D. thesis,
  \bibinfo{school}{The University of British Columbia (Vancouver)}
  (\bibinfo{year}{2009}).

\bibitem[{\citenamefont{Smyth and Winters}(2003)}]{Smyth-Winters-2003}
\bibinfo{author}{\bibfnamefont{W.~D.} \bibnamefont{Smyth}} \bibnamefont{and}
  \bibinfo{author}{\bibfnamefont{K.~B.} \bibnamefont{Winters}},
  \bibinfo{journal}{J. Phys. Oceanogr.} \textbf{\bibinfo{volume}{33}},
  \bibinfo{pages}{694} (\bibinfo{year}{2003}).

\bibitem[{\citenamefont{Alexakis}(2005)}]{Alexakis-2005}
\bibinfo{author}{\bibfnamefont{A.}~\bibnamefont{Alexakis}},
  \bibinfo{journal}{Phys. Fluids} \textbf{\bibinfo{volume}{17}},
  \bibinfo{pages}{084103} (\bibinfo{year}{2005}).

\bibitem[{\citenamefont{Alexakis}(2007)}]{Alexakis-2007}
\bibinfo{author}{\bibfnamefont{A.}~\bibnamefont{Alexakis}},
  \bibinfo{journal}{Phys. Fluids} \textbf{\bibinfo{volume}{19}},
  \bibinfo{pages}{054105} (\bibinfo{year}{2007}).

\bibitem[{\citenamefont{Rayleigh}(1880)}]{Rayleigh-1880}
\bibinfo{author}{\bibfnamefont{L.}~\bibnamefont{Rayleigh}},
  \bibinfo{journal}{Proc. London Math. Soc.} \textbf{\bibinfo{volume}{11}},
  \bibinfo{pages}{57} (\bibinfo{year}{1880}).

\bibitem[{\citenamefont{Baines and Mitsudera}(1994)}]{Baines-Mitsudera-1994}
\bibinfo{author}{\bibfnamefont{P.}~\bibnamefont{Baines}} \bibnamefont{and}
  \bibinfo{author}{\bibfnamefont{H.}~\bibnamefont{Mitsudera}},
  \bibinfo{journal}{J. Fluid Mech.} \textbf{\bibinfo{volume}{276}}
  (\bibinfo{year}{1994}).

\bibitem[{\citenamefont{Caulfield}(1994)}]{Caulfield-1994}
\bibinfo{author}{\bibfnamefont{C.~P.} \bibnamefont{Caulfield}},
  \bibinfo{journal}{J. Fluid Mech.} \textbf{\bibinfo{volume}{258}},
  \bibinfo{pages}{255} (\bibinfo{year}{1994}).

\bibitem[{\citenamefont{Haigh and Lawrence}(1999)}]{Haigh-Lawrence-1999}
\bibinfo{author}{\bibfnamefont{S.~P.} \bibnamefont{Haigh}} \bibnamefont{and}
  \bibinfo{author}{\bibfnamefont{G.~A.} \bibnamefont{Lawrence}},
  \bibinfo{journal}{Phys. Fluids} \textbf{\bibinfo{volume}{11}},
  \bibinfo{pages}{1459} (\bibinfo{year}{1999}).

\bibitem[{\citenamefont{Harnik et~al.}(2008)\citenamefont{Harnik, Heifetz,
  Umurhan, and Lott}}]{Harnik-etal-2008}
\bibinfo{author}{\bibfnamefont{N.}~\bibnamefont{Harnik}},
  \bibinfo{author}{\bibfnamefont{E.}~\bibnamefont{Heifetz}},
  \bibinfo{author}{\bibfnamefont{O.~M.} \bibnamefont{Umurhan}},
  \bibnamefont{and} \bibinfo{author}{\bibfnamefont{F.}~\bibnamefont{Lott}},
  \bibinfo{journal}{J. Atmos. Sci.} \textbf{\bibinfo{volume}{65}},
  \bibinfo{pages}{2615} (\bibinfo{year}{2008}).

\bibitem[{\citenamefont{Rabinovich et~al.}(2011)\citenamefont{Rabinovich,
  Umurhan, Harnik, Lott, and Heifetz}}]{Rabinovich-etal-2011}
\bibinfo{author}{\bibfnamefont{A.}~\bibnamefont{Rabinovich}},
  \bibinfo{author}{\bibfnamefont{O.~M.} \bibnamefont{Umurhan}},
  \bibinfo{author}{\bibfnamefont{N.}~\bibnamefont{Harnik}},
  \bibinfo{author}{\bibfnamefont{F.}~\bibnamefont{Lott}}, \bibnamefont{and}
  \bibinfo{author}{\bibfnamefont{E.}~\bibnamefont{Heifetz}},
  \bibinfo{journal}{J. Fluid Mech.} \textbf{\bibinfo{volume}{670}},
  \bibinfo{pages}{301} (\bibinfo{year}{2011}).

\bibitem[{\citenamefont{Carpenter et~al.}(2010)\citenamefont{Carpenter,
  Balmforth, and Lawrence}}]{Carpenter-etal-2010}
\bibinfo{author}{\bibfnamefont{J.~R.} \bibnamefont{Carpenter}},
  \bibinfo{author}{\bibfnamefont{N.~J.} \bibnamefont{Balmforth}},
  \bibnamefont{and} \bibinfo{author}{\bibfnamefont{G.~A.}
  \bibnamefont{Lawrence}}, \bibinfo{journal}{Phys. Fluids}
  \textbf{\bibinfo{volume}{22}}, \bibinfo{pages}{054104}
  (\bibinfo{year}{2010}).

\bibitem[{\citenamefont{Bretherton}(1966)}]{Bretherton-1966}
\bibinfo{author}{\bibfnamefont{F.~P.} \bibnamefont{Bretherton}},
  \bibinfo{journal}{Q. J. R. Meteorol. Soc.} \textbf{\bibinfo{volume}{92}},
  \bibinfo{pages}{335} (\bibinfo{year}{1966}).

\bibitem[{\citenamefont{Sakai}(1989)}]{Sakai-1989}
\bibinfo{author}{\bibfnamefont{S.}~\bibnamefont{Sakai}}, \bibinfo{journal}{J.
  Fluid Mech.} \textbf{\bibinfo{volume}{202}}, \bibinfo{pages}{149}
  (\bibinfo{year}{1989}).

\bibitem[{\citenamefont{Heifetz and Methven}(2005)}]{Heifetz-Methven-2005}
\bibinfo{author}{\bibfnamefont{E.}~\bibnamefont{Heifetz}} \bibnamefont{and}
  \bibinfo{author}{\bibfnamefont{J.}~\bibnamefont{Methven}},
  \bibinfo{journal}{Phys. Fluids} \textbf{\bibinfo{volume}{17}},
  \bibinfo{pages}{064107} (\bibinfo{year}{2005}).

\bibitem[{\citenamefont{{Bakas} and Ioannou}(2009)}]{Bakas-Ioannou-2009}
\bibinfo{author}{\bibfnamefont{N.~A.} \bibnamefont{{Bakas}}} \bibnamefont{and}
  \bibinfo{author}{\bibfnamefont{P.~J.} \bibnamefont{Ioannou}},
  \bibinfo{journal}{Phys. Fluids} \textbf{\bibinfo{volume}{21}},
  \bibinfo{pages}{024102} (\bibinfo{year}{2009}).

\bibitem[{\citenamefont{Goldreich et~al.}(1986)\citenamefont{Goldreich,
  Goodman, and Narayan}}]{Goldreich-etal-1986}
\bibinfo{author}{\bibfnamefont{P.}~\bibnamefont{Goldreich}},
  \bibinfo{author}{\bibfnamefont{J.}~\bibnamefont{Goodman}}, \bibnamefont{and}
  \bibinfo{author}{\bibfnamefont{R.}~\bibnamefont{Narayan}},
  \bibinfo{journal}{Mon. Not. R. Astron. Soc.} \textbf{\bibinfo{volume}{221}},
  \bibinfo{pages}{339} (\bibinfo{year}{1986}).

\bibitem[{\citenamefont{Narayan et~al.}(1987)\citenamefont{Narayan, Goldreich,
  and Goodman}}]{Narayan-etal-1987}
\bibinfo{author}{\bibfnamefont{R.}~\bibnamefont{Narayan}},
  \bibinfo{author}{\bibfnamefont{P.}~\bibnamefont{Goldreich}},
  \bibnamefont{and} \bibinfo{author}{\bibfnamefont{J.}~\bibnamefont{Goodman}},
  \bibinfo{journal}{Mon. Not. R. Astron. Soc.} \textbf{\bibinfo{volume}{228}},
  \bibinfo{pages}{1} (\bibinfo{year}{1987}).

\bibitem[{\citenamefont{Umurhan}(2010)}]{Umurhan-2010}
\bibinfo{author}{\bibfnamefont{O.~M.} \bibnamefont{Umurhan}},
  \bibinfo{journal}{A\&A} \textbf{\bibinfo{volume}{521}} (\bibinfo{year}{2010}).

\bibitem[{\citenamefont{Farrell and Ioannou}(1996)}]{Farrell-Ioannou-1996a}
\bibinfo{author}{\bibfnamefont{B.~F.} \bibnamefont{Farrell}} \bibnamefont{and}
  \bibinfo{author}{\bibfnamefont{P.~J.} \bibnamefont{Ioannou}},
  \bibinfo{journal}{J. Atmos. Sci.} \textbf{\bibinfo{volume}{53}},
  \bibinfo{pages}{2025} (\bibinfo{year}{1996}).

\bibitem[{\citenamefont{Schmid and Henningson}(2001)}]{Schmid-Henningson-2001}
\bibinfo{author}{\bibfnamefont{P.~J.} \bibnamefont{Schmid}} \bibnamefont{and}
  \bibinfo{author}{\bibfnamefont{D.~S.} \bibnamefont{Henningson}},
  \emph{\bibinfo{title}{Stability and Transition in Shear Flows}}
  (\bibinfo{publisher}{Springer, New York}, \bibinfo{year}{2001}).

\bibitem[{\citenamefont{Kelvin}(1887)}]{Kelvin-1887b}
\bibinfo{author}{\bibfnamefont{L.}~\bibnamefont{Kelvin}},
  \bibinfo{journal}{Phil. Mag. (5)} \textbf{\bibinfo{volume}{24}},
  \bibinfo{pages}{188} (\bibinfo{year}{1887}).

\bibitem[{\citenamefont{Orr}(1907)}]{Orr-1907}
\bibinfo{author}{\bibfnamefont{W.~M.} \bibnamefont{Orr}},
  \bibinfo{journal}{Proc.~Roy. Irish Acad.} \textbf{\bibinfo{volume}{27}},
  \bibinfo{pages}{9} (\bibinfo{year}{1907}).

\bibitem[{\citenamefont{Phillips}(1966)}]{Phillips-1966}
\bibinfo{author}{\bibfnamefont{O.~M.} \bibnamefont{Phillips}},
  \emph{\bibinfo{title}{The Dynamics of the Upper Ocean}}
  (\bibinfo{publisher}{Cambridge University Press, Cambridge},
  \bibinfo{year}{1966}), chap.~\bibinfo{chapter}{5}, pp.
  \bibinfo{pages}{178--184}.

\bibitem[{\citenamefont{Hartmann}(1975)}]{Hartmann-1975}
\bibinfo{author}{\bibfnamefont{R.~J.} \bibnamefont{Hartmann}},
  \bibinfo{journal}{J. Fluid Mech.} \textbf{\bibinfo{volume}{71}},
  \bibinfo{pages}{89} (\bibinfo{year}{1975}).

\bibitem[{\citenamefont{Farrell and
  Ioannou}(1993{\natexlab{a}})}]{Farrell-Ioannou-1993c}
\bibinfo{author}{\bibfnamefont{B.~F.} \bibnamefont{Farrell}} \bibnamefont{and}
  \bibinfo{author}{\bibfnamefont{P.~J.} \bibnamefont{Ioannou}},
  \bibinfo{journal}{J. Atmos. Sci.} \textbf{\bibinfo{volume}{50}},
  \bibinfo{pages}{2201} (\bibinfo{year}{1993}{\natexlab{a}}).

\bibitem[{\citenamefont{Farrell and
  Ioannou}(1993{\natexlab{b}})}]{Farrell-Ioannou-1993b}
\bibinfo{author}{\bibfnamefont{B.~F.} \bibnamefont{Farrell}} \bibnamefont{and}
  \bibinfo{author}{\bibfnamefont{P.~J.} \bibnamefont{Ioannou}},
  \bibinfo{journal}{Physics of Fluids} \textbf{\bibinfo{volume}{5}},
  \bibinfo{pages}{2298} (\bibinfo{year}{1993}{\natexlab{b}}).

\bibitem[{\citenamefont{Heifetz et~al.}(1999)\citenamefont{Heifetz, Bishop, and
  Alpert}}]{Heifetz-etal-1999}
\bibinfo{author}{\bibfnamefont{E.}~\bibnamefont{Heifetz}},
  \bibinfo{author}{\bibfnamefont{C.~H.} \bibnamefont{Bishop}},
  \bibnamefont{and} \bibinfo{author}{\bibfnamefont{P.}~\bibnamefont{Alpert}},
  \bibinfo{journal}{Q. J. R. Meteorol. Soc.} \textbf{\bibinfo{volume}{125}},
  \bibinfo{pages}{2835} (\bibinfo{year}{1999}).

\bibitem[{\citenamefont{Lott}(1997)}]{Lott-1997}
\bibinfo{author}{\bibfnamefont{F.}~\bibnamefont{Lott}}, \bibinfo{journal}{Q. J.
  R. Meteorol. Soc.} \textbf{\bibinfo{volume}{123}}, \bibinfo{pages}{1603}
  (\bibinfo{year}{1997}).

\bibitem[{\citenamefont{{Bakas} and Ioannou}(2007)}]{Bakas-Ioannou-2007}
\bibinfo{author}{\bibfnamefont{N.~A.} \bibnamefont{{Bakas}}} \bibnamefont{and}
  \bibinfo{author}{\bibfnamefont{P.~J.} \bibnamefont{Ioannou}},
  \bibinfo{journal}{J. Atmos. Sci.} \textbf{\bibinfo{volume}{64}},
  \bibinfo{pages}{1509} (\bibinfo{year}{2007}).

\bibitem[{\citenamefont{Lott et~al.}(2010)\citenamefont{Lott, Plougonven, and
  Vanneste}}]{Lott-etal-2010}
\bibinfo{author}{\bibfnamefont{F.}~\bibnamefont{Lott}},
  \bibinfo{author}{\bibfnamefont{R.}~\bibnamefont{Plougonven}},
  \bibnamefont{and} \bibinfo{author}{\bibfnamefont{J.}~\bibnamefont{Vanneste}},
  \bibinfo{journal}{J. Atmos. Sci.} \textbf{\bibinfo{volume}{67}},
  \bibinfo{pages}{157} (\bibinfo{year}{2010}).

\bibitem[{\citenamefont{Farrell}(1988{\natexlab{a}})}]{Farrell-1988b}
\bibinfo{author}{\bibfnamefont{B.~F.} \bibnamefont{Farrell}},
  \bibinfo{journal}{J. Atmos. Sci.} \textbf{\bibinfo{volume}{45}},
  \bibinfo{pages}{163} (\bibinfo{year}{1988}{\natexlab{a}}).

\bibitem[{\citenamefont{Farrell}(1989)}]{Farrell-1989}
\bibinfo{author}{\bibfnamefont{B.~F.} \bibnamefont{Farrell}},
  \bibinfo{journal}{J. Atmos. Sci.} \textbf{\bibinfo{volume}{46}},
  \bibinfo{pages}{1193} (\bibinfo{year}{1989}).

\bibitem[{\citenamefont{Farrell and
  Ioannou}(1993{\natexlab{c}})}]{Farrell-Ioannou-1993d}
\bibinfo{author}{\bibfnamefont{B.~F.} \bibnamefont{Farrell}} \bibnamefont{and}
  \bibinfo{author}{\bibfnamefont{P.~J.} \bibnamefont{Ioannou}},
  \bibinfo{journal}{J. Atmos. Sci.} \textbf{\bibinfo{volume}{50}},
  \bibinfo{pages}{4044} (\bibinfo{year}{1993}{\natexlab{c}}).

\bibitem[{\citenamefont{Farrell and Ioannou}(1994)}]{Farrell-Ioannou-1994b}
\bibinfo{author}{\bibfnamefont{B.~F.} \bibnamefont{Farrell}} \bibnamefont{and}
  \bibinfo{author}{\bibfnamefont{P.~J.} \bibnamefont{Ioannou}},
  \bibinfo{journal}{Phys. Rev. Lett.} \textbf{\bibinfo{volume}{72}},
  \bibinfo{pages}{1118} (\bibinfo{year}{1994}).

\bibitem[{\citenamefont{Umurhan and Heifetz}(2007)}]{Umurhan-Heifetz-2007}
\bibinfo{author}{\bibfnamefont{O.~M.} \bibnamefont{Umurhan}} \bibnamefont{and}
  \bibinfo{author}{\bibfnamefont{E.}~\bibnamefont{Heifetz}},
  \bibinfo{journal}{Phys. Fluids} \textbf{\bibinfo{volume}{19}},
  \bibinfo{pages}{064102} (\bibinfo{year}{2007}).

\bibitem[{\citenamefont{Hazel}(1972)}]{Hazel-1972}
\bibinfo{author}{\bibfnamefont{P.}~\bibnamefont{Hazel}}, \bibinfo{journal}{J.
  Fluid Mech.} \textbf{\bibinfo{volume}{51}}, \bibinfo{pages}{39}
  (\bibinfo{year}{1972}).

\bibitem[{\citenamefont{Smyth and Peltier}(1989)}]{Smyth-Peltier-1989}
\bibinfo{author}{\bibfnamefont{W.~D.} \bibnamefont{Smyth}} \bibnamefont{and}
  \bibinfo{author}{\bibfnamefont{W.~R.} \bibnamefont{Peltier}},
  \bibinfo{journal}{J. Atmos. Sci.} \textbf{\bibinfo{volume}{46}},
  \bibinfo{pages}{3698} (\bibinfo{year}{1989}).

\bibitem[{\citenamefont{Smyth and Peltier}(1990)}]{Smyth-Peltier-1990}
\bibinfo{author}{\bibfnamefont{W.~D.} \bibnamefont{Smyth}} \bibnamefont{and}
  \bibinfo{author}{\bibfnamefont{W.~R.} \bibnamefont{Peltier}},
  \bibinfo{journal}{Geophys. Astrophys. Fluid Dyn.}
  \textbf{\bibinfo{volume}{52}}, \bibinfo{pages}{249} (\bibinfo{year}{1990}).

\bibitem[{\citenamefont{Howard and Maslowe}(1973)}]{Howard-Maslowe-1973}
\bibinfo{author}{\bibfnamefont{L.~N.} \bibnamefont{Howard}} \bibnamefont{and}
  \bibinfo{author}{\bibfnamefont{S.~A.} \bibnamefont{Maslowe}},
  \bibinfo{journal}{Boundary Layer Meteorol.} \textbf{\bibinfo{volume}{4}},
  \bibinfo{pages}{511} (\bibinfo{year}{1973}).

\bibitem[{\citenamefont{Farrell}(1988{\natexlab{b}})}]{Farrell-1988a}
\bibinfo{author}{\bibfnamefont{B.~F.} \bibnamefont{Farrell}},
  \bibinfo{journal}{Physics of Fluids} \textbf{\bibinfo{volume}{31}},
  \bibinfo{pages}{2093} (\bibinfo{year}{1988}{\natexlab{b}}).

\bibitem[{\citenamefont{Hill}(1995)}]{Hill-95}
\bibinfo{author}{\bibfnamefont{D.~C.} \bibnamefont{Hill}}, \bibinfo{journal}{J.
  Fluid Mech.} \textbf{\bibinfo{volume}{292}}, \bibinfo{pages}{183}
  (\bibinfo{year}{1995}).

\bibitem[{\citenamefont{Marquet et~al.}(2009)\citenamefont{Marquet, Lombardi,
  Chomaz, Sipp, and Jacquin}}]{Chomaz-etal-2009}
\bibinfo{author}{\bibfnamefont{O.}~\bibnamefont{Marquet}},
  \bibinfo{author}{\bibfnamefont{M.}~\bibnamefont{Lombardi}},
  \bibinfo{author}{\bibfnamefont{J.-M.} \bibnamefont{Chomaz}},
  \bibinfo{author}{\bibfnamefont{D.}~\bibnamefont{Sipp}}, \bibnamefont{and}
  \bibinfo{author}{\bibfnamefont{D.}~\bibnamefont{Jacquin}},
  \bibinfo{journal}{J. Fluid Mech.} \textbf{\bibinfo{volume}{622}},
  \bibinfo{pages}{1} (\bibinfo{year}{2009}).

\bibitem[{\citenamefont{Ioannou}(1995)}]{Ioannou-1995}
\bibinfo{author}{\bibfnamefont{P.~J.} \bibnamefont{Ioannou}},
  \bibinfo{journal}{J. Atmos. Sci.} \textbf{\bibinfo{volume}{52}},
  \bibinfo{pages}{1155} (\bibinfo{year}{1995}).

\end{thebibliography}

\begin{figure*}
\includegraphics{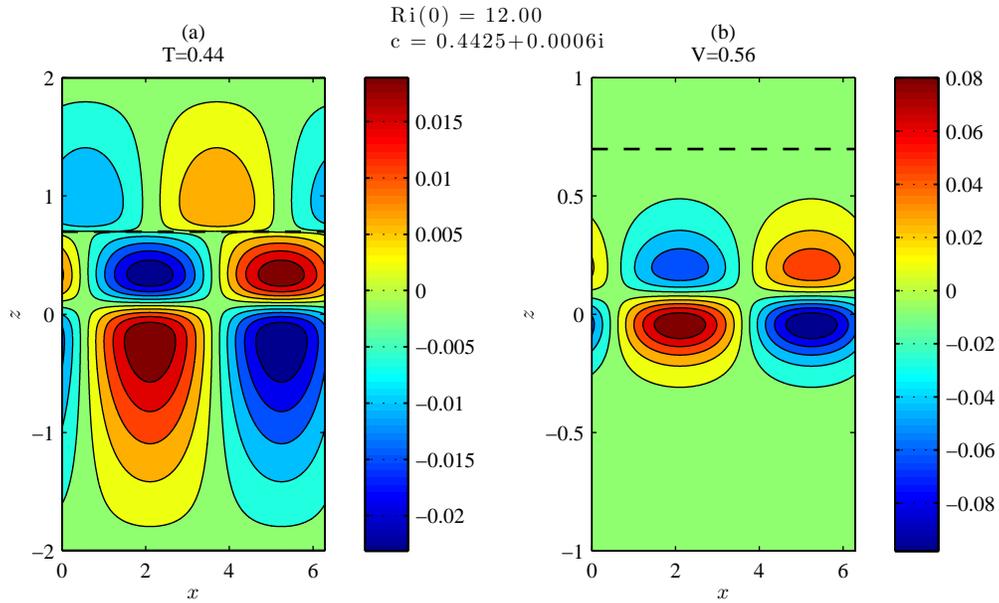}
\caption{\label{fig:Holmboe_mode2}(Color online) Structure of the streamfunction (panel (a)) and density (panel (b)) perturbation fields of unstable prograde H2 mode with growth rate $kc_i = 0.0006$ and phase speed $c_r=0.4425$. Critical layer at $z_c=0.698$ (dashed). For perturbations with $k=1$, and base flow with $R=3$ and $\Ri(0)=12.00$.}
\end{figure*}

\begin{figure*}
\includegraphics{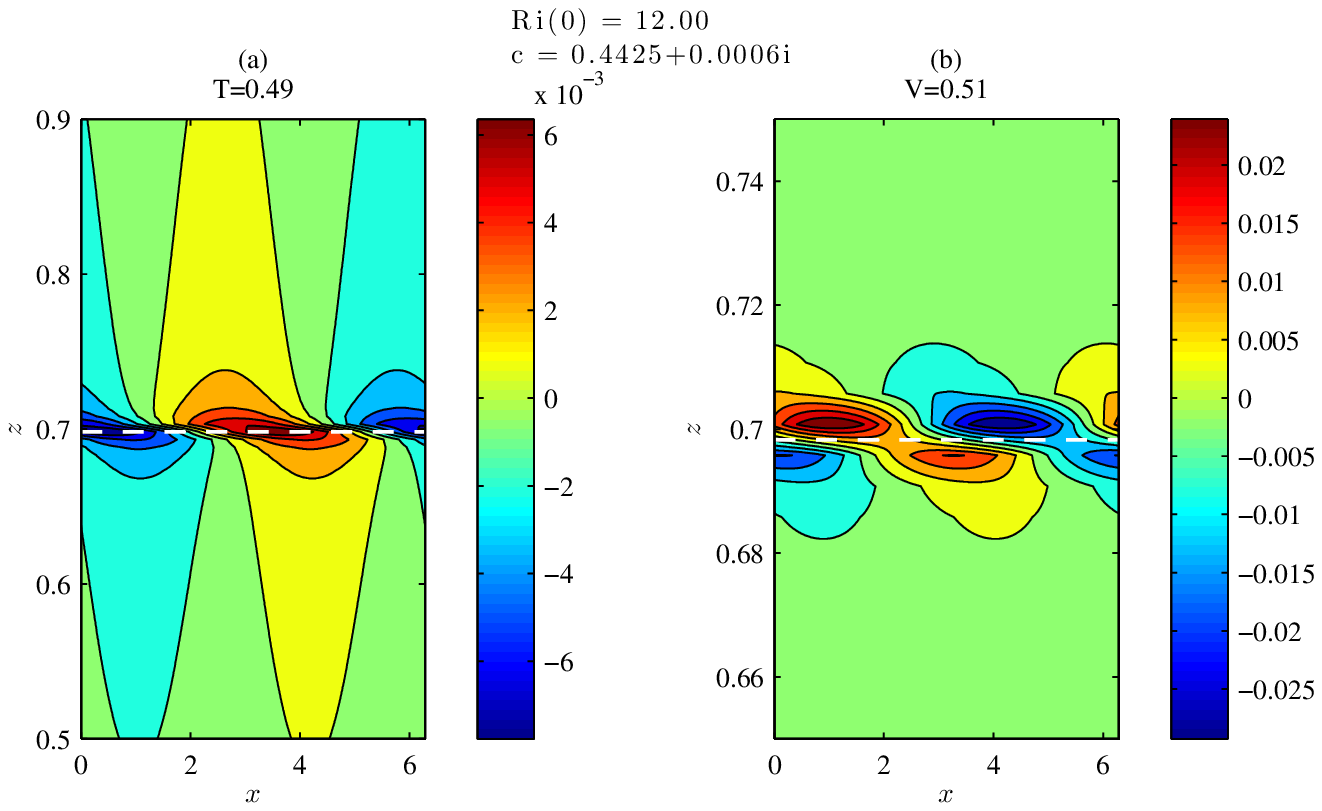}
\caption{\label{fig:Holmboe_mode2_adjoints}(Color online) Structure of  the streamfunction and density perturbation fields of the adjoint of the unstable prograde H2 mode of  FIG.~\ref{fig:Holmboe_mode2}. The adjoint is centered at the steering level, $z_c= 0.698$, of the mode.} 
\end{figure*}

\begin{figure*}
\includegraphics{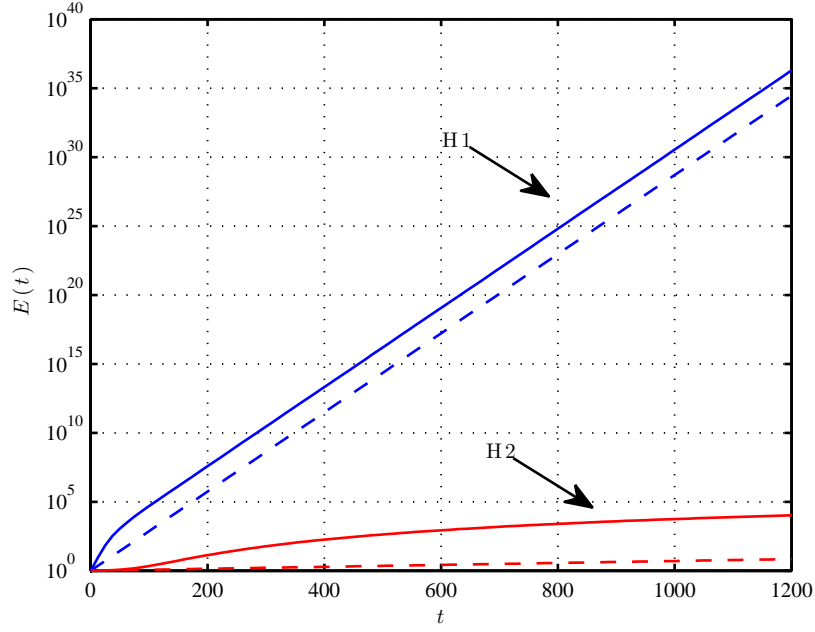}
\caption{\label{fig:H1H2adjt}
(Color online) Time evolution of the perturbation energy when the adjoint  of the H1 mode shown in FIG.~\ref{fig:Holmboe_mode1_adjoints}
and the  H2 mode shown in FIG.~\ref{fig:Holmboe_mode2_adjoints} are introduced at $t=0$ (solid); also shown is the energy evolution if the H1 mode and H2 mode are introduced at $t=0$ (dashed). The adjoint excites the corresponding modes optimally. 
The adjoint of the H1 mode excites the mode at an amplitude a factor $69$ greater in energy than an initial condition consisting of the H1 mode itself. 
The H2 adjoint excites the mode at an amplitude a factor $7921$ greater in energy.
Parameters are $k=1$, $R=3$  and for the H1 mode the stratification is $\Ri(0)=0.65$, while for the H2 mode $\Ri(0)=12$.}
\end{figure*}


\clearpage


\end{document}